%% file: Draft.tex
\documentclass[
    prl, twocolumn, superscriptaddress, nofootinbib, amsmath, amssymb,
    aps, floatfix, preprintnumbers
]{revtex4-2}
\usepackage[utf8]{inputenc}
\usepackage{setspace}
\usepackage{graphicx}
\usepackage{tikz}
\definecolor{CiteBlue}{RGB}{45,52,151}
\usepackage[
    colorlinks=true,
    linkcolor=CiteBlue,
    urlcolor=CiteBlue,
    citecolor=CiteBlue
]{hyperref}
\usepackage{orcidlink}
\usepackage{amsmath}

\newcommand{\ignore}[1]{}

\begin{document}

\title{Analytical Fluxes from Generic Schwarzschild Geodesics}

\author{Majed Khalaf\,\orcidlink{0000-0001-5537-9992}}
\affiliation{Racah Institute of Physics, Hebrew University of Jerusalem, Jerusalem 91904, Israel}

\author{Chris Kavanagh\,\orcidlink{0000-0002-2874-9780}}
\affiliation{School of Mathematics and Statistics, University College Dublin, Belfield D04 N2E5, Dublin 4, Ireland}

\author{Ofri Telem\,\orcidlink{0000-0002-3120-0975}}
\affiliation{Racah Institute of Physics, Hebrew University of Jerusalem, Jerusalem 91904, Israel}

\date\today

\begin{abstract}\ignorespaces{}
   % We present a method to analytically compute gravitational wave fluxes from Schwarzschild geodesics with generic eccentricities. Our method is based on a systematic expansion of the Fourier coefficients of the emitted radiation, using Chebyshev polynomials. This allows us to effectively express these Fourier coefficients as a sum of Keplerian-like Fourier coefficients, known from our previous work on the Quantum Spectral Method. We check our results to 15PN and reproduce the overall flux from a geodesic with $p=12, e=0.5$ to $10^{-5}$ accuracy. 
   We present an analytic method for computing gravitational-wave fluxes from bound Schwarzschild geodesics with arbitrary eccentricity. Our approach systematically expands the Fourier coefficients of the emitted radiation in a Chebyshev basis, allowing them to be reduced to sums of Keplerian-like Fourier coefficients previously derived in the Quantum Spectral Method. Because the construction does not rely on a small-eccentricity expansion, it applies to a broad range of bound eccentric orbits. As an illustration, we implement the method using a $15$PN-expanded input and find that it reproduces the total flux for the case $(p,e)=(12.5,0.5)$ to relative accuracy $10^{-5}$, while for the stronger-field case $(p,e)=(10,0.8)$ it yields weighted mode-by-mode errors below $10^{-6}$ for the selected dominant modes analyzed. These results provide an analytic route to frequency-domain flux calculations relevant to EMRI modeling. 
\end{abstract}

\maketitle
\textbf{Introduction.}
Over the past decade, the detection of gravitational wave signals by terrestrial observatories have transformed our understanding of mergers involving black holes and neutron stars \cite{LIGOScientific:2016aoc,LIGOScientific:2017vwq,LIGOScientific:2020ibl}. The upcoming space-based detector, the Laser Interferometer Space Antenna (LISA) \cite{LISA:2017pwj,LISA:2022yao,LISA:2024hlh}, will enable the first observations of extreme mass-ratio inspirals (EMRIs), events in which a stellar mass compact object inspirals into a much more massive black hole, with typical mass ratios on the range $10^{-4}-10^{-6}$. Unlike the short inspiral phase in most mergers observed by ground-based detectors, the inspiral phase of EMRIs can be on the order of years \cite{Babak:2017tow}.

Maintaining accuracy of gravitational wave templates over such long timescales represents a significant modeling challenge, and is predominantly tackled using the self-force approach (see, e.g., \cite{Barack:2018yvs,Pound:2021qin} and references therein). The self-force approach uses the small mass-ratio to describe the phase evolution of EMRI waveforms using a post-adiabatic (PA) expansion, with complete models requiring the leading order adiabatic (0PA) and first subleading (1PA) terms \cite{Hinderer:2008dm}. At the 1PA frontier, recent work has demonstrated the power of combining numerical self-force information with analytic weak-field post-Newtonian results to generate hybrid models which extend non-spinning 1PA models without requiring computationally expensive numerical self-force computations \cite{Mathews:2025txc,Honet:2025gge,Burke:2023lno}. This motivates the generation of further analytic models including all relevant orbital dynamics such as eccentricity and spin dependencies. Notably however, the current frontier of analytic self-force modeling has been limited to a low eccentricity expansion \cite{Munna:2020iju, Munna:2023wce,Sago:2015rpa,Sago:2026gxb}, with closed form or resummed expressions only available for particular terms \cite{Munna:2020som} . 

In the post-adiabatic framework, computations traditionally utilize a two-pronged approach: a first computationally expensive \textit{off-line} step, in which, for example, energy and angular momentum fluxes are computed throughout the EMRI parameter space \cite{Drasco:2005kz,Skoupy:2023lih}; and a highly efficient \textit{on-line} step, in which an inspiralling trajectory and its associated waveform can be extracted in milliseconds \cite{Chua:2020stf,Katz:2021yft}. In this paper we consider the off-line flux computation stage in the frequency domain \cite{Hughes:2021exa}%,Osburn:2014hoa}  
(see also \cite{Harms:2014dqa,Sundararajan_2007} for time-domain computations). Frequency-domain computations involve the projection of time-dependent observables into a basis of Fourier elements with respect to a given geodesic in the following manner (restricting for this work to the case of equatorial motion). For any function $f(r)$, its $n,m$ Fourier element is defined as
\begin{align} \label{SchFour}
    \left\{f\right\}^{nm}_i & \equiv\frac{1}{\pi}\int_{r_{p}}^{r_{a}}f\left(r\right)\,\cos[ A^i_{nm} (r)]\,t'_i(r)\,dr\,,
\end{align}
where $A^i_{nm}(r) = \omega^i_{nm} t_i(r)- m\,\varphi_{i}(r)$, $r_p$ ($r_a$) is the periapsis (apoapsis), and $\omega^i_{nm}\equiv \Omega^r_{i} n + \Omega^\varphi_{i} m$ is the frequency of the $(n,m)$ harmonic of the geodesic. The label $i$ indicates the system of interest -- in our case geodesic motion in Schwarzschild spacetime, but later we will expand it using Fourier elements for Keplerian motion. 

In this work, we present an algorithm that allows us to analytically compute any Fourier coefficient associated with geodesic motion in Schwarzschild, with a straightforward extention to equatorial motion in Kerr spacetime. Using simple algebraic manipulations and integration-by-parts, any Schwarzschild Fourier element $\left\{f\right\}^{nm}_{\mathrm S}$ can be expressed as a series of Keplerian Fourier elements denoted by $\left\{\mathcal{K}\left[f\right]/[1+\alpha/(Er)]\right\}^{nm}_{\mathrm K}$, where $\mathcal{K}$ is a linear integral operator defined below, and $\mathcal{K}\left[f\right]$ is expanded into a Laurent-like series of monomials $r^j,\,j\in\mathbb{Z}$. The parameters of the auxiliary Keplerian system are chosen so that the Keplerian turning-point radii coincide with their Schwarzschild counterparts. The Keplerian Fourier coefficients for every monomial, $\left\{r^j/[1+\alpha/(Er)]\right\}^{nm}_{\mathrm K}$, were computed in previous work \cite{Khalaf:2023ozy,Khalaf:2025jpt} using the Quantum Spectral Method. The bulk of this work focuses on an efficient expansion of $\mathcal{K}\left[f\right]$ into a Laurent-like series. This series should be uniformly convergent in the range $r_{\rm{min}}\leq r \leq r_{\rm{max}}$ where the Fourier integral is performed. For this purpose, a purely post-Newtonian expansion centered at $r\rightarrow\infty$ is of little use. To this end, we make use of an expansion in terms of \textit{Chebyshev polynomials}, well adapted for uniform convergence over a given interval, in particular for the periodic functions we encounter in our computation. Ultimately, our expansion is a hybrid of PN and Chebyshev expansions, in which periodic functions in the $r_{\rm{min}}\leq r \leq r_{\rm{max}}$ range are expanded in Chebyshev polynomials, while their well-behaved prefactors are PN expanded in $r^{-1}$. Importantly, our algorithm does not involve an expansion in eccentricity, and so it is valid both for circular and for highly eccentric geodesics. For the purpose of the flux computations of this paper, our seed Fourier kernel $f$ is based on a post-Newtonian (PN) expansion of the Green's function for the Teukolsky equation to 15PN order, generated using the Black Hole Perturbation Toolkit (BHPT) \cite{BHPToolkit, Castillo:2024isq}.

Using our algorithm, we are able to compute the energy and angular momentum fluxes from Schwarzschild geodesics of general eccentricity. As a proof-of-principle, we compute the full energy flux from a $p=12.5,\,e=0.5$ geodesic, as well as from selected modes of a $p=10,\,e=0.8$ geodesic. We benchmark our computation with the numerical results of the BHPT and find an $\mathcal{O}(10^{-5})$ error in the total flux for $p=12.5,\,e=0.5$, and at most a $\mathcal{O}(10^{-7})$ error per mode for $p=10,\,e=0.8$. Though we did not optimize for performance in this proof-of-principle, we note that our \textit{Mathematica} implementation runs an $\mathcal{O}(1)$ faster than its numerical counterpart using the BHPT.

\textbf{Gravitational flux computations.} 
The leading order radiated energy and angular momentum flux from an EMRI is encapsulated in the projection of the gravitational perturbation $h_{\mu\nu}$ into the Weyl curvature scalar $\psi_4$. In Schwarzschild, $\psi_4$ decomposes as
\begin{eqnarray}
\psi_4=r^{-4} \sum_{n\ell m} R_{n\ell m}(r){}_{-2} Y_{\ell m}(\theta) e^{-i \omega_{nm}^{\mathrm S} t+i m \phi}\,,
\end{eqnarray}
where the ${}_{s}Y_{\ell m}$ are the spin-weighted spherical harmonics, and $\omega_{nm }^{\mathrm S}=\Omega^r_{\mathrm S} n+\Omega^\varphi_{\mathrm S} m$ are the frequencies of the geodesic sourcing the gravitational waves (GWs). The radial functions $R_{n\ell m}(r)$ are solutions of the inhomogeneous radial Teukolsky equation in Schwarzschild spacetime with a point particle source \cite{Teukolsky:1972my,Teukolsky:1974yv}. They are expressed as a linear combination of two homogeneous solutions of the Teukolsky equation: $R_{\ell m }^{\rm up}(r,\omega)$ which is outgoing at infinity, and $R_{\ell m }^{\rm in}(r,\omega)$ which is incoming at the horizon. $R_{\ell m }^{\rm up/in}(r,\omega)$ are well known in a low-frequency expansion \cite{Mano:1996mf,Mano:1996vt}, have been expanded to high PN orders \cite{Fujita:2012cm, Kavanagh:2016idg} and can be computed within the \texttt{Teukolsky} package of the BHPT \cite{BHPToolkit}. Explicitly, $R_{n\ell m}(r)$ is given as
\begin{eqnarray}
R_{n\ell m}(r)=Z_{n\ell m}^{+}(r) R_{\ell m }^{\rm up}(r,\omega_{nm}^{\mathrm S})+Z_{\ell m  n}^{-}(r) R_{\ell m}^{\mathrm{in}}(r,\omega_{nm}^{\mathrm S}).
\end{eqnarray}
 Asymptotically, the coefficients $Z_{n\ell m}^{\pm}(r)$ are constants and can be written as Fourier elements of the form 
\begin{eqnarray}\label{eq:Z}
&&Z_{n\ell m}^{+}=\left\{\mathcal{D}_r R_{\ell m}^{\mathrm{in}}\right\}^{nm}_{\mathrm S}~~,~~Z_{n\ell m}^{-}=\left\{\mathcal{D}_rR_{\ell m}^{\mathrm{up}}\right\}^{nm}_{\mathrm S}\,.
\end{eqnarray}
Here $\mathcal{D}_r$ is a 2nd order integro-differential operator in $r$ given in the Supplemental Material (SM), and the $\mathrm{S}$ indicates a Fourier decomposition with respect to geodesic motion in Schwarzschild. Once $Z_{n\ell m}^{\pm}$ is computed, the energy and angular momentum fluxes at infinity are given by the well known expressions
\begin{eqnarray}
\left\langle\frac{d E}{d t}\right\rangle^{\infty}&=&\sum_{n\ell m} \frac{1}{4 \pi \left(\omega_{nm}^{\mathrm S}\right)^2}\left|Z_{n\ell m}^{+}\right|^2\nonumber\\[5pt]\left\langle\frac{d L_z}{d t}\right\rangle^{\infty}&=&\sum_{n \ell m} \frac{m}{4 \pi \left(\omega_{nm}^{\mathrm S}\right)^3}\left|Z_{n \ell m}^{+}\right|^2 \,,
\end{eqnarray}
with a similar expression for the fluxes at the horizon. In the rest of this paper, we will devise an analytical method for computing \eqref{eq:Z}, by expanding in Keplerian Fourier elements.

\textbf{Schwarzschild and Kepler EOM.} 
The equations of motion (EOM) for Schwarzschild and relativistic Keplerian geodesic motion are well known, and we present them as ODEs in the independent variable $r$. Due to the spherical symmetry, the motion in both cases is planar, % Without loss of generality, we consider trajectories in the XY plane.
and conveniently parametrized by the functions $t_i(r),\,\phi_i(r)$. Here $i={\rm S}\,(\rm K)$ denotes the Schwarzschild (relativistic Keplerian) case. A given orbit is uniquely specified (up to an initial phase) by its angular momentum and orbital energy. We denote the angular momentum by $L_i$, while we always consider $E_{\rm S}=E_{\mathrm K}\equiv E$. Throughout this manuscript, we work in units such that $c=GM=1$.

The EOM in Schwarzschild are then given by
\begin{align} \label{EOMSch}
\frac{dt_{\mathrm S}}{dr}=\frac{Er^{4}}{\Delta(r)\sqrt{U_{\mathrm S}^{r}\left(r\right)}}\,\,,\,\, \frac{d\varphi_{\mathrm S}}{dr}=\frac{L_{\mathrm S}}{\sqrt{U_{\mathrm S}^{r}\left(r\right)}}\,,
\end{align}
where $\Delta (r) = r\left(r - 2 \ignore{GM}\right)$ and $U_{\mathrm S}^{r}\left(r\right)=r^{4}\left[E^{2}-\frac{\Delta\left(r\right)}{r^{2}}\left(\frac{L^{2}_{\mathrm S}}{r^{2}}+\mu^{2}\right)\right]$ where $\mu$ is the mass of the probe. The three nonzero roots of $U^r_{\mathrm S}(r)$ are $r_{a}\equiv\frac{\ignore{G M} p}{1-e},\,r_{p}\equiv\frac{\ignore{G M} p}{1+e}$, and $r_h\equiv\frac{2\ignore{GM} p}{p-4}$, where $p$ and $e$ are the semilatus rectum and eccentricity, respectively, $E=\mu \sqrt{\frac{(p-2)^2-4 e^2}{\left(p-e^2-3\right) p}},\,L_{\mathrm S}=\frac{\ignore{GM}\mu p}{\sqrt{p-e^2-3}}$.
For bounded motion $0\leq e<1$, we have $0<r_h<r_{p}\leq r\leq r_{a}$.

For special relativistic Keplerian motion we have
\begin{align} \label{EOMKep}
\frac{dt_{\mathrm K}}{dr}=\frac{E+\frac{\alpha}{r}}{\sqrt{U_{\mathrm K}^{r}\left(r\right)}}\,\,,\,\, \frac{d\varphi_{\mathrm K}}{dr}=\frac{L_{\mathrm K}}{r^2\sqrt{U_{\mathrm K}^{r}\left(r\right)}}\,,
\end{align}
with $U_{\mathrm K}^{r}\left(r\right)=E^{2}-\mu^{2}+\frac{2 E \alpha}{r}-\frac{L_{\mathrm K}^2-\alpha^2}{r^{2}}$. The roots of $U_{\mathrm K}^{r}\left(r\right)$ are $r_{\mathrm{max}/\mathrm{min}}=E\alpha \frac{1\pm\sqrt{1-E^{-2} \alpha^{-2}\left(\mu^2-E^2\right)\left(L_{\mathrm K}^2-\alpha^2\right)}}{\mu^2-E^2}$. Below we present our algorithm for the reduction of Schwarzschild Fourier elements to a series of Keplerian ones. Crucially, this reduction is well behaved when we choose Keplerian parameters that guarantee that $r_{\mathrm{min}}=r_{p}$ and $r_{\mathrm{ max}}=r_{a}$. This happens when
\begin{align} \label{KLKepcond}
    &\frac{\alpha}{\ignore{G M} \mu}=(p-4) \sqrt{\frac{p}{\left((p-2)^2-4 e^2\right)
   \left(-e^2+p-3\right)}}\,,\nonumber\\
   &\frac{L_{\mathrm K}}{\alpha} = \sqrt{\frac{p(p-3)-4 e^2}{p-4}}\,.
\end{align}
The corresponding fundamental frequencies are given by $\Omega^r_i=\frac{2\pi}{T^r_i},\,\Omega^\varphi_i=\Omega^r_i \Delta\varphi_i /\pi$ as well as
 \begin{eqnarray}
&&T^r_i=2\int^{r_{a}}_{r_{p}}t'_i(r)\,dr~~,~~\Delta\varphi_i=\int^{r_{a}}_{r_{p}}\varphi'_i(r)\,dr\,.
\end{eqnarray}
\indent \textbf{Reduction to Keplerian Fourier elements.} 
For our choice \eqref{KLKepcond} of Keplerian parameters, we can reduce Schwarzschild Fourier elements to Keplerian ones in the following manner. First, we define the linear operator $\mathcal{K}_{nm}$ acting on any function $f(r)$ as:
\begin{eqnarray}
\mathcal{K}_{nm}[f]\equiv \mathcal{A}_{nm}(r)f\left(r\right)+\mathcal{B}_{nm}(r)\int^r_{r_{p}}\mathcal{C}_{nm}(r')f(r')dr'\,
\end{eqnarray}
where
\begin{eqnarray}
\mathcal{A}_{nm}(r)&=&\frac{r^2}{\Delta(r)\sqrt{1-\frac{r_h}{r}}}\cos\left[ \Delta A_{nm}\left(r\right)\right]\,\nonumber\\[5pt]
\mathcal{B}_{nm}(r)&=&\frac{E}{\sqrt{\mu^2-E^2}}\left[\omega^K_{nm}\left(1+\frac{\alpha}{Er}\right) -\frac{m L_{\mathrm K}}{E r^{2}}\right]\nonumber\\[5pt]
\mathcal{C}_{nm}(r)&=&\frac{r^3\sin\left(\Delta A_{nm}(r)\right)}{\Delta(r)\sqrt{1-\frac{r_h}{r}}\sqrt{\left(r-r_{p}\right)\left(r_{a}-r\right)}}\,,
\end{eqnarray}
and $\Delta A_{n m}\equiv A^{\mathrm S}_{n m}-A^{\mathrm K}_{n m}$.
It is then straightforward to check the following relation:
\begin{eqnarray}\label{eq:schtokep}
\left\{f\right\}^{nm}_{\mathrm S} = \left\{\frac{\mathcal{K}_{nm}[f]}{1+\frac{\alpha}{E r}}\right\}^{nm}_{\mathrm K}\,.
\end{eqnarray}
This is the fundamental relation allowing us to compute Schwarzschild Fourier coefficients from Keplerian ones. Importantly, this relation does not involve an expansion in eccentricity, and so it allows us to perform analytical computation for Schwarzschild geodesics of \textit{generic eccentricity}. 
Below we will present a robust and efficient algorithm for expanding $\mathcal{K}_{nm}\left[f\right]$ as a \textit{Laurent-like} series in $r$, namely a sum of monomials $r^j,\,j\in\mathbb{Z}$ approximating $\mathcal{K}_{nm}[f]$ for $r_{p}\leq r\leq r_{a}$. Once this is done, \eqref{eq:schtokep} can be analytically computed using the Keplerian Fourier coefficient formula \cite{Khalaf:2023ozy,Khalaf:2025jpt}

\begin{align} \label{KepElem}
& \left\{\frac{r^j}{1+\frac{\alpha}{E r}}\right\}^{nm}_{\mathrm K}= \rho(-1)^{n} \left(\frac{\ignore{GM} p e}{2\left(1-e^2\right)}\right)^j\nonumber\\
& \times \left(\frac{1+\sqrt{1-e^2}}{e}\right)^{j-n} (j-\tilde{m}-n+2)_n\nonumber \\
&\times\sum _{k=0}^{\infty} \frac{ \left(\frac{1-\sqrt{1-e^2}}{e}\right)^{2k}   (-j-\tilde{m}-1)_{k}  (-j+\tilde{m}+n-1)_{k} }{k! \Gamma (k+n+1)}\nonumber\\
&\times\,_1F_1\left(-k;j-k+\tilde{m}+2;-(n+\tilde{m})\rho\right) \nonumber\\
&\times \, _1F_1\left(-k-n;j-k-\tilde{m}-n+2;(n+\tilde{m})\rho\right)\,.
\end{align}
Here $\rho\left(p,e\right)\equiv\frac{\left(1+\sqrt{1-e^2}\right)\left((p-2)^2-4 e^2\right)}{2 p \left(p-e^2-3\right)}$, and $\tilde{m}\equiv m \left[1+(p-4)/\left(4 e^2+3p-p^2\right)\right]^{-1/2}$. 

\textbf{Laurent-like series for $\mathcal{K}_{nm}[f]$.} Our expansion of $\mathcal{K}_{nm}[f]$ is a hybrid of a PN expansion and an expansion in terms of Chebyshev/Gegenbauer polynomials (see also \cite{Whittall:2025dqn} for a recent application of Gegenbauer polynomials in numerical self-force computations). The latter are crucial to ensure convergence over the entire interval $r_{p}\leq r\leq r_{a}$. 

In this section, we assume that $f(r)$ is already in the form of a Laurent-like series, namely is a sum of monomials $r^j$ with $j\in \mathbb{Z}$. As we will see below, this requires some pre-processing in the case of Schwarzschild flux coefficients. Once $f(r)$ is Laurent-like, the first step in forming our Laurent-like series for $\mathcal{K}_{nm}[f]$ is the Chebyshev expansion of $\cos[\Delta A_{nm}\left(r\right)]$ and $\sin[\Delta A_{nm}\left(r\right)]$. We do this in two steps; first, we expand them in terms of their argument. Then, we substitute the general form of $\Delta A_{nm}\left(r\right)$, and finally, we Chebyshev-expand the result. We write the expansions of $\cos(\Delta A_{nm})$ and $\sin(\Delta A_{nm})$ symbolically as
\begin{eqnarray}\label{eq:cossinCheb}
\cos(\Delta A_{nm})&=&J_0(a)+2\sum^{\infty}_{i=1}\,(-1)^i\,J_{2i}(a)\,\,T_{2i}(\Delta A_{nm}/a)\nonumber\\[5pt]
\sin(\Delta A_{nm})&=&2\sum^{\infty}_{i=0}(-1)^i\,J_{2i+1}(a)\,\,T_{2i+1}(\Delta A_{nm}/a)\,,~~~~~~
\end{eqnarray}
where the $T_k(x)$ are Chebyshev polynomials of the first kind and $J_i(a)$ are Bessel functions evaluated at an arbitrary value $a$. In practice, we cut-off the expansions at $N_{\cos} = 5,\,N_{\sin} = 6$. While $a$ is formally arbitrary, the choice of $a$ matters greatly when truncating the sums, with our choice of $a=\Delta A_{nm}\left(\frac{r_p + r_a}{2}\right)$ being near-optimal.

Next, we substitute into \eqref{eq:cossinCheb} the general form of $\Delta A_{n m}(r)$. As we show in the SM, it is given by 
\begin{eqnarray} \label{DeltaASKform}
    &&\Delta A_{n m}(r) ={\sqrt{\left(r-r_{p}\right)\left(r_{a}-r\right)}}\times \nonumber\\
    &&\left[ \chi_1^{(1)}\left(r\right)\, n + \chi_1^{(2)}\left(r\right)\, m+ \left(\chi_2^{(1)} n + \chi_2^{(2)} m\right)g(r)\right],\nonumber\\
    \end{eqnarray}
    where
  \begin{eqnarray} \label{eqg}
    &&g(r)=\frac{\arctan\left[\frac{\sqrt{\left(r-r_{p}\right)\left(r_{a}-r\right)}}{r+\sqrt{r_{p}r_{a}}}\right]}{{\sqrt{\left(r-r_{p}\right)\left(r_{a}-r\right)}}}\,
  \end{eqnarray}  
and $\chi^{(1,2)}_{1}(r)$ are functions of $r$ that depend on $p,e$. They are easily expanded in PN perturbation theory, where  they assume a polynomial form in $r^{-1}$. Note that these functions are well behaved throughout the domain $r_p\leq r\leq r_p$, and so a simple expansion in $r^{-1}$ suffices. The $\chi^{\left(1,2\right)}_2$'s are \textit{independent} of $r$. Substituting \eqref{DeltaASKform} into \eqref{eq:cossinCheb} and noting that $T_{2i}(x)$ only involves even powers while $T_{2i+1}(x)$ only involves odd powers, we see that all terms in $\mathcal{A}_{mn}(r),\,\mathcal{C}_{mn}(r)$ are now of the form $L(r)\,[g(r)]^k$ where $L(r)$ is Laurent-like. We can expand $g(r)$ in Chebyshev polynomials as
\begin{eqnarray} \label{gcheb}
  &&g(r)=\frac{2}{r_{a}-r_{p}} \sum_{i=0}^{\infty}  \frac{(-1)^iq^{i+1}}{i+1}U_{i}\left(x\right)\,\nonumber\\[5pt]
  &&x\equiv\frac{2 r-r_{a}-r_{p}}{r_{a}-r_{p}}~~,~~q\equiv\frac{\sqrt{r_a}-\sqrt{r_p}}{\sqrt{r_a}+\sqrt{r_p}}\,,
\end{eqnarray}  
where the $U_i(x)$ are Chebyshev polynomials of the second kind. The Chebyshev expansion for $[g(r)]^k$ follows iteratively, using $U_i(x) U_j(x)=\sum_{s=0}^{\min (i, j)} U_{i+j-2 s}(x)$.

\begin{figure}[ht]
  \centering
  
  \includegraphics[width=0.45\textwidth]{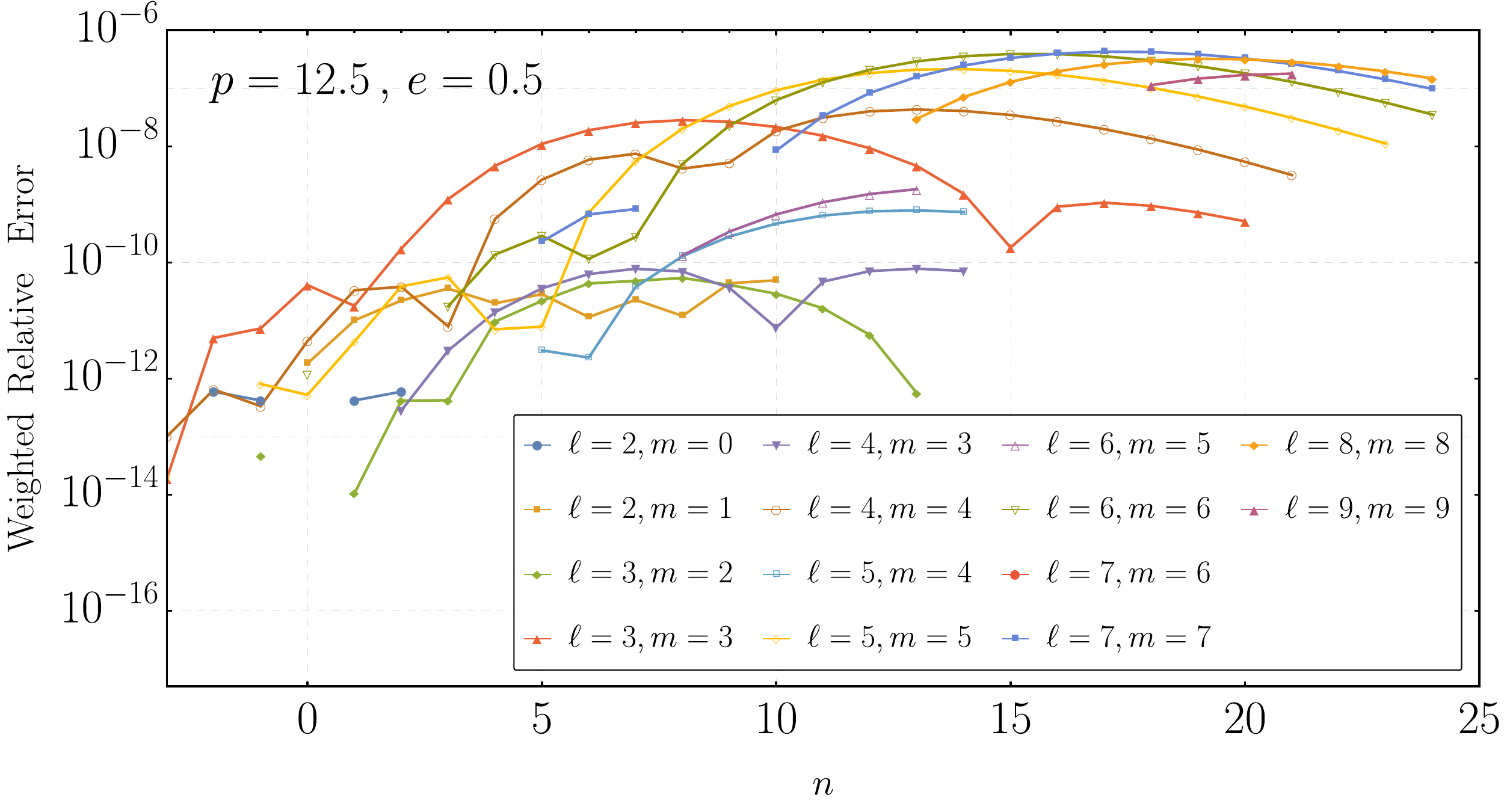}
  \includegraphics[width=0.45\textwidth]{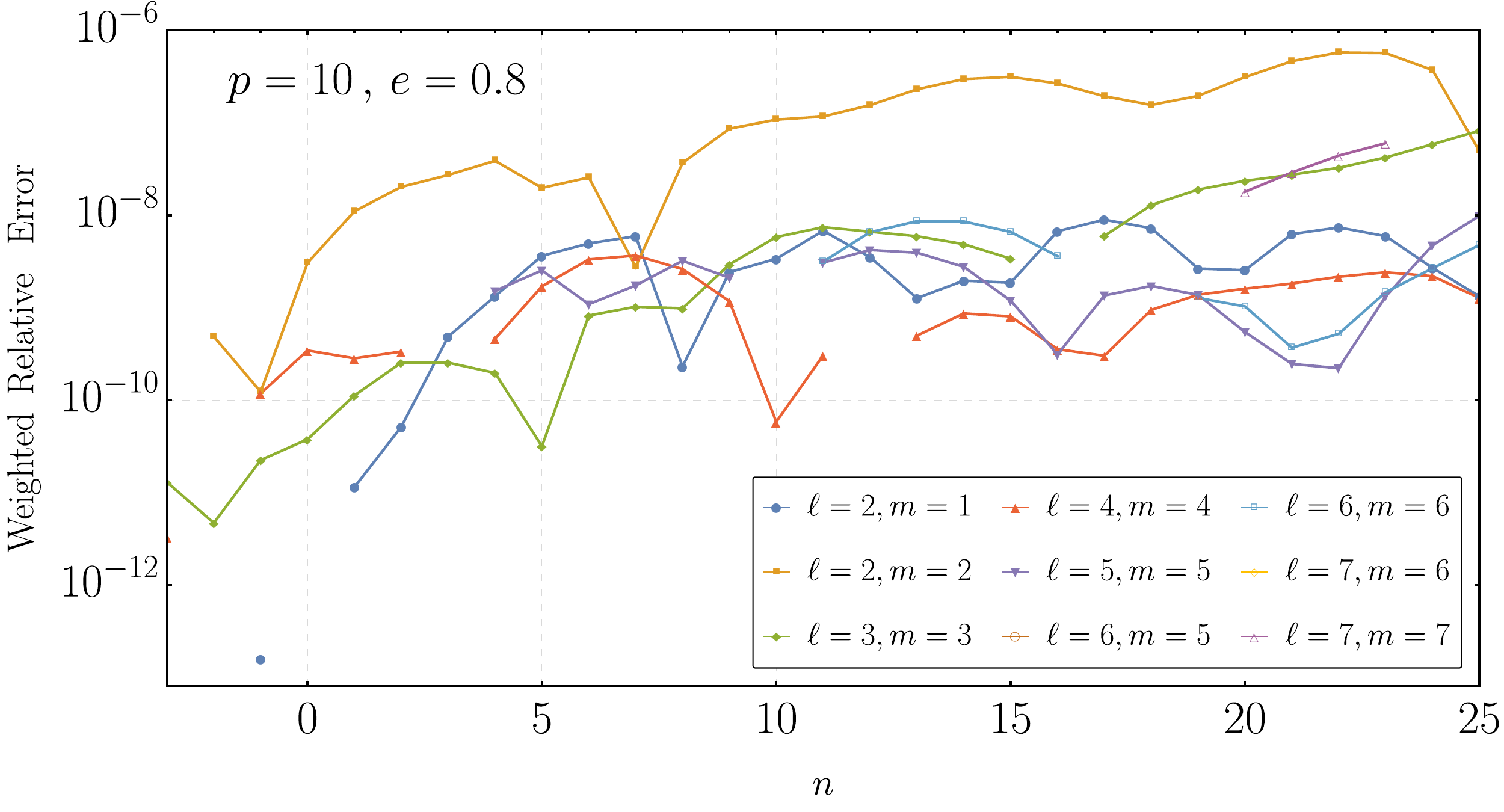}
  \caption{Radiation Flux per $(l,m)$ mode, emitted from a Schwarzschild geodesic. We plot the weighted relative error between our analytical result, computed with a 15PN Green's function, and the numerically computed fluxes from the Black Hole Perturbation Toolkit. The horizontal axis is the mode number $n$. Top: $p=12.5, e=0.5$, Bottom: $p=10, e=0.8$. In the top (bottom) panel, we only include modes whose relative contribution to the total flux exceeds $10^{-6}$ ($10^{-7}$). 
  }
  \label{fig:fluxes}
\end{figure}

\textbf{Application to flux coefficients}. We apply our reduction algorithm to compute the Fourier elements $Z^{+}_{n\ell m}$ defined\footnote{The $Z^{-}_{n\ell m}$ can be evaluated in a similar way to compute horizon fluxes, but we did not do this explicitly in this work.} in \eqref{eq:Z}. As a first step, we use a 15PN-expanded form of $R^{\mathrm{in}}_{\ell m}$ computed with the BHPT \cite{BHPToolkit}. The expression $\mathcal{D}_r R^{\mathrm{in}}_{\ell m}$ contains Laurent-like terms in $r$ as well as $\log^k\left(r\right),\,k\in \mathbb{N}$ and $\sqrt{(r-r_p)(r_a - r)}$. We deal with the latter using integration-by-parts (see SM). For the $\log^k$ term, we employ the Gegenbauer expansion
\begin{align} \label{eq:gegen}
\log^k(r) &= \sum_{i=0}^{\infty} \sum_{j=0}^{k}
\binom{k}{j}
\left( 2 \log\left( \frac{\sqrt{r_p} + \sqrt{r_a}}{2} \right) \right)^{k - j}\nonumber\\
&\times (-1)^{i + j} \, q^i \,
\left. \frac{\partial^j}{\partial \lambda^j} C_i^{(\lambda)}(x) \right|_{\lambda = 0}\, .
\end{align}  
In practice, we truncate the infinite sums at $N_{\log} = 15$. The result is a Laurent-type expansion for $\mathcal{D}_r R^{\mathrm{in}}_{\ell m}$ of the form $\sum_{i=-\infty}^\infty \frac{a_i}{r^i}$. Finally, we used \eqref{eq:schtokep} to compute $Z^{\mathrm{in}}_{n\ell m}$, expanding $\mathcal{K}[\mathcal{D}_rR^{\mathrm{in}}_{\ell m}]$ with up to 15 terms in its Chebyshev expansion. We benchmarked our results versus the numerically computed $\dot{E}_{n\ell m}\equiv |Z^{\mathrm{in}}_{n\ell m}|^2/(4\pi\omega^2_{mn})$ as a function of $n$ for different values of $l,m$. Our results are depicted in Figure~\ref{fig:fluxes} for $p=12.5,\,e=0.5$ and for $p=10,\,e=0.8$. The vertical axis of these plots is the weighted relative error (WRE) defined as
\begin{eqnarray}
{\rm WRE}=\frac{\dot{E}^{\rm BHPT}_{n\ell m}}{\dot{E}^{\rm BHPT}_{tot}}\left|\frac{\dot{E}^{\rm analytical}_{n\ell m}-\dot{E}^{\rm BHPT}_{n\ell m}}{\dot{E}^{\rm BHPT}_{n\ell m}}\right|\,.
\end{eqnarray} 
This weighting is natural as modes with lesser accuracy also tend to be less significant. Figure~\ref{fig:fluxes} was produced using $\chi^{(1,2)}_{1}$ expanded through $15\mathrm{PN}$ order in the top panel and $20\mathrm{PN}$ order in the bottom panel. The expansion of $\left[g(r)\right]^k$ was truncated at (Chebyshev) order $16$ in the top panel, and at orders $25$ ($k=1$) and $27$ ($k>1$) in the bottom panel. Finally, in the SM, we illustrate the main features of our computations with an explicit toy calculation of the flux coefficients at $2\mathrm{PN}$ order.
% To produce Figure~\ref{fig:fluxes}, we used $\chi^{(1,2)}_{1}$ expanded through $15$PN order in the top panel, and $20$PN in the bottom panel. The cutoffs of the expansion for $\left[g(r)\right]^k$ was $16$ in the top panel, whereas for the bottom panel it was $25$ for $k=1$ and $27$ for higher $k$.

\textbf{Conclusions.}
In this work we presented a robust algorithm for the analytical computation of Schwarzschild Fourier elements by expanding them in a series of Keplerian Fourier elements. The expansion utilizes Chebyshev/Gegenbauer polynomials for uniform convergence in $r_{\rm{min}}\leq r\leq r_{\rm{max}}$, which is crucial for reliable results. We applied our algorithm for the computation of gravitational wave fluxes from a point mass on an eccentric Schwarzschild geodesic, relevant for 0PA self-force computations, and benchmarked our results against the numerical computations using the BHPT.

Our work establishes a new pipeline for the flux computations at the heart of modern adiabatic self-force computations. It enables both a nontrivial analytical cross-check of existing numerical computations and a robust way to extend them to highly eccentric orbits. 

Our success in analytically computing the fluxes from Schwarzschild geodesics calls for several future research directions. First and foremost, our methods should be directly generalizable to Kerr fluxes from generic geodesics -- even eccentric and non-equatorial ones. Secondly, in \cite{Khalaf:2025jpt} we showed how relativistic Keplerian Fourier elements, computed using \eqref{KepElem}, can be analytically continued to the unbound regime and benchmarked against post-Minkowski/post-Newtonian computations. Using the algorithm outlined in this manuscript, this continuation could be extended to the Schwarzschild case as well. Finally, the analytical insight gained in our computations could be utilized at the post-Adiabatic order as well, in order to extend the state-of-the-art 2SF computations to eccentric orbits.

\textbf{Acknowledgments.} We thank Leor Barack, Riccardo Gonzo, Donal O'Connell, and Adam Pound for useful discussions. MK is grateful to the Azrieli Foundation for the award of
an Azrieli Fellowship. MK is supported in part by grants No. 2022713 and  2024091 from the US-Israel BSF/NSF, grant No. 2023711 from the US-Israel BSF, and by an ERC STG grant (``Light-Dark,'' grant No. 101040019). OT is supported by the ISF grant No. 3533/24, by the NSF-BSF grant No. 2022713, and by the BSF grant No. 2024169.
This project has received funding from the European Research Council (ERC) under the European Union’s Horizon Europe research and innovation programme (grant agreement No. 101040019). Views and opinions expressed are however those of the author(s) only and do not necessarily reflect those of the European Union. The European Union cannot be held responsible for them.  CK acknowledges support from Research Ireland under
Grant number 21/PATH-S/9610. CK and MK are grateful for support from the Nordita program 'Amplitudes, Strong-Field Gravity and Resummation' during the final stages of this work. 
%%%%%%%%%%%%%%%%%%%%%%%%%%%%%%%%%

\bibliography{References}

\include{SM.tex}
\end{document}

%% file: SM.tex
\onecolumngrid
% \appendix

\setcounter{equation}{0}
\setcounter{figure}{0}
\setcounter{table}{0}
\setcounter{page}{1}
\renewcommand{\theequation}{S\arabic{equation}}
\renewcommand{\thefigure}{S\arabic{figure}}
\renewcommand{\thepage}{S\arabic{page}}
\makeatletter

\begin{center}
\textbf{\large Supplemental Material for: \textit{Analytical Fluxes from Generic Schwarzschild Geodesics}}
\end{center}
In this Supplemental Material, we provide further details on definitions, derivations and computations omitted from the main text. In particular, in Section~A we derive the general form of $\Delta A_{nm}$; in Section~B we define $\mathcal{D}_r$ and explain how to treat the square-root term; and in Section~C we present a toy calculation illustrating the main features of the algorithm introduced in this work. Scaling this calculation yields the benchmarks shown in Fig.~\ref{fig:fluxes} of the main text.
\section{A.~General form of $\Delta A_{n m}(r)$ }
In this section, we study the form of $\Delta A_{nm}$ and elaborate on~\eqref{DeltaASKform} of the main text. To begin, note that

% \[
% A^{i}_{nm}(r)=\omega^{i}_{nm}t_{i}\left(r\right)-m\,\varphi_{i}\left(r\right)=\left(n+\frac{\Omega^{\varphi}_{i}}{\Omega^{r}_{i}}m\right)\Omega^{r}_{i}t_{i}\left(r\right)-m\,\varphi_{i}\left(r\right)
% \]
\begin{align}
\Delta A_{nm}\left(r\right)&=A^{{\rm S}}_{nm}-A^{{\rm K}}_{nm}=\left(n+\frac{\Omega^{\varphi}_{{\rm S}}}{\Omega^{r}_{{\rm S}}}m\right)\Omega^{r}_{{\rm S}}t_{{\rm S}}\left(r\right)-\left(n+\frac{\Omega^{\varphi}_{{\rm K}}}{\Omega^{r}_{{\rm K}}}m\right)\Omega^{r}_{{\rm K}}t_{{\rm K}}\left(r\right)-m\,\left(\varphi_{{\rm S}}\left(r\right)-\varphi_{{\rm K}}\left(r\right)\right)\nonumber\\
&=\left(n+\frac{\Omega^{\varphi}_{{\rm S}}}{\Omega^{r}_{{\rm S}}}m\right)\left(\Omega^{r}_{{\rm S}}t_{{\rm S}}\left(r\right)-\Omega^{r}_{{\rm K}}t_{{\rm K}}\left(r\right)\right)+\left(\frac{\Omega^{\varphi}_{{\rm S}}}{\Omega^{r}_{{\rm S}}}-\frac{\Omega^{\varphi}_{{\rm K}}}{\Omega^{r}_{{\rm K}}}\right)m\Omega^{r}_{{\rm K}}t_{{\rm K}}\left(r\right)-m\,\left(\varphi_{{\rm S}}\left(r\right)-\varphi_{{\rm K}}\left(r\right)\right)\nonumber\\
&\equiv\left(n+\frac{\Omega^{\varphi}_{{\rm S}}}{\Omega^{r}_{{\rm S}}}m\right)\Delta q^{r}\left(r\right)+\left(\frac{\Omega^{\varphi}_{{\rm S}}}{\Omega^{r}_{{\rm S}}}-\frac{\Omega^{\varphi}_{{\rm K}}}{\Omega^{r}_{{\rm K}}}\right)m\,q^{r}_{{\rm K}}\left(r\right)-m\,\Delta\varphi\left(r\right)\,,
\end{align}
where $\Delta q^r \equiv q^r_{\rm S} - q^r_{\rm K}$, with $q^r_{i} \equiv \Omega^r_i t_i$ being the action-angle in Schwarzschild coordinates. Explicitly, by using~\eqref{EOMSch} and~\eqref{EOMKep} of the main text,
\begin{equation}\label{Appeq:qrKexp}
q^{r}_{{\rm K}}\left(r\right)=\Omega^{r}_{{\rm Kep}}\int^{r}_{r_{p}}\frac{E+\frac{\alpha}{r'}}{\sqrt{U^{r}_{{\rm Kep}}\left(r'\right)}}\,\,dr'=\frac{\left(e^{2}-1\right)(2e-p+2)(2e+p-2)}{p^{2}\left(e^{2}-p+3\right)}\sqrt{(r_{a}-r)(r-r_{p})}+2\arctan\left(\sqrt{\frac{r-r_{p}}{r_{a}-r}}\right)\,,
\end{equation}
and
\begin{equation} \label{appeq:qsandchi0}
q^{r}_{{\rm S}}\left(r\right)=\Omega^{r}_{{\rm Sch}}\int^{r}_{r_{p}}\frac{\chi^{\left(0\right)}\left(r'\right)r'}{\sqrt{\left(r'-r_{p}\right)\left(r_{a}-r'\right)}}\,\,dr',\quad \chi^{\left(0\right)}\left(r\right)\equiv\frac{E}{\sqrt{\mu^{2}-E^{2}}}\frac{1}{\left(1-\frac{2}{r}\right)\sqrt{1-\frac{r_{h}}{r}}}\equiv\sum^{\infty}_{k=0}\beta_{k}\left(p,e\right)r^{-k}\,.
\end{equation}
In principle, one can calculate $q^r_{\rm S}$ by using the expansion of $\chi^{\left(0\right)}\left(r\right)$ and
the explicit integrals
\begin{subequations} \label{appeq: integrals}
 \begin{align}
     \int^{r}_{r_{p}}\frac{r'}{\sqrt{\left(r'-r_{p}\right)\left(r_{a}-r'\right)}}\,\,dr' &= 2r_{+}\arctan\left(\sqrt{\frac{r-r_{p}}{r_{a}-r}}\right)-\sqrt{\left(r-r_{p}\right)\left(r_{a}-r\right)}\,,\\
     \int^{r}_{r_{p}}\frac{1}{\sqrt{\left(r'-r_{p}\right)\left(r_{a}-r'\right)}}\,\,dr' &= 2\arctan\left(\sqrt{\frac{r-r_{p}}{r_{a}-r}}\right)\,,\\
     \int^{r}_{r_{p}}\frac{1}{r'^{\ell}\sqrt{\left(r'-r_{p}\right)\left(r_{a}-r'\right)}}\,\,dr' &= \frac{\left(-1\right)^{\ell-1}}{\left(\ell-1\right)!}\frac{\partial^{\ell-1}}{\partial r^{\ell-1}_{+}}\left[\frac{2\arctan\left(\sqrt{\frac{r_{+}+r_{-}}{r_{+}-r_{-}}}\sqrt{\frac{r-r_{p}}{r_{a}-r}}\right)}{\sqrt{r^{2}_{+}-r^{2}_{-}}}\right]\,,
 \end{align}
\end{subequations}
where $\ell$ is a positive integer, and $r_{\pm}\equiv\frac{r_a\pm r_p}{2}$. However, since we are interested
in $\Delta q^{r}$, we can simplify the calculation as follows. From
the form of the integrals above, it is straightforward to see that $q^{r}_{{\rm S}}$ is made of three terms; a function
expanded in $r^{-1}$ multiplied by $\sqrt{\left(r-r_{p}\right)\left(r_{a}-r\right)}$,
an $r-$independent coefficient multiplied by $\arctan\left(\sqrt{\frac{r_{a}}{r_{p}}}\sqrt{\frac{r-r_{p}}{r_{a}-r}}\right)$,
and finally an analogous piece for $\arctan\left(\sqrt{\frac{r-r_{p}}{r_{a}-r}}\right)$.
Combined with~\eqref{Appeq:qrKexp}, the form of $\Delta q^{r}$ remains
unchanged. By virtue of the matching conditions, $\Delta q^{r}\left(r_{p}\right)=\Delta q^{r}\left(r_{a}\right)=0$ must hold. As a consequence, the coefficients of $\arctan\left(\sqrt{\frac{r_{a}}{r_{p}}}\sqrt{\frac{r-r_{p}}{r_{a}-r}}\right)$
and $\arctan\left(\sqrt{\frac{r-r_{p}}{r_{a}-r}}\right)$ must be
opposite to each other, and we can write
\begin{equation}\label{appeq:deltaqr}
\Delta q^{r}\left(r\right)=\chi^{\left(1\right)}_{1}\left(r\right)\sqrt{\left(r-r_{p}\right)\left(r_{a}-r\right)}+\chi^{\left(1\right)}_{2}\arctan\left[\frac{\sqrt{\left(r-r_{p}\right)\left(r_{a}-r\right)}}{r+\sqrt{r_{p}r_{a}}}\right]\,,\quad \chi^{\left(1\right)}_{2}\equiv2\left(1-\Omega^{r}_{{\rm Sch}}r_{+}\beta_{0}-\Omega^{r}_{{\rm Sch}}\beta_{1}\right)\,,
\end{equation}
and $\chi^{\left(1\right)}_{1}$ is a function of $r$ that admits
a PN expansion. Using the integrals~\eqref{appeq: integrals}, one obtains
% \[
% \chi^{\left(1\right)}_{1}\left(r\right)=\Omega^{r}_{{\rm Sch}}\left[\frac{\left(1-e^{2}\right)\left(2+2e-p\right)\left(2+2e+p\right)}{p^{2}\left(3+e^{2}-p\right)}-\beta_{0}+\sum^{\infty}_{k=3}\beta_{k}\left(p,e\right)\frac{\left(-1\right)^{k}}{\left(k-2\right)!}\frac{\partial^{k-2}}{\partial r^{k-2}_{+}}\left[\frac{2\arctan\left(\sqrt{\frac{r_{+}+r_{-}}{r_{+}-r_{-}}}\sqrt{\frac{r-r_{p}}{r_{a}-r}}\right)}{\sqrt{r^{2}_{+}-r^{2}_{-}}}\right]\right]
% \]
\begin{equation} \label{appeq: chi11}
\chi^{\left(1\right)}_{1}\left(r\right)=\frac{\left(1-e^{2}\right)\left(2+2e-p\right)\left(2e+p-2\right)}{p^{2}\left(3+e^{2}-p\right)}-\Omega^{r}_{{\rm Sch}}\beta_{0}+\Omega^{r}_{{\rm Sch}}\sum^{\infty}_{k=3}\beta_{k}\left(p,e\right)\frac{\left(\sqrt{1-e^{2}}\right)^{k}}{p^{k}}P^{\left(1\right)}_{k-2}\left(r\right)\,,
\end{equation}
where
\begin{equation}
P^{\left(1\right)}_{k}\left(r\right)\equiv\left.\sum^{k-1}_{s=0}\frac{(-1)^{s}}{(1+s)!}\left(\sqrt{1-e^{2}}\right)^{s+1}P_{k-1-s}\!\left(\frac{1}{\sqrt{1-e^{2}}}\right)\frac{d^{s}}{dx^{s}}\left[\frac{\sqrt{\frac{1-e^{2}}{x^{2}-e^{2}}}}{x-1+(1-e^{2})\,\frac{r}{p}}\right]\right|_{x=1}\,,
\end{equation}
where $P_n\left(x\right)$ is a Legendre polynomial of order $n$.

We now move on to the remaining terms in $\Delta A_{nm}$. From~\eqref{EOMSch} and~\eqref{EOMKep} of the main text, 

\begin{align}
\Delta\varphi\left(r\right) & =\frac{L}{\sqrt{\mu^{2}-E^{2}}}\int^{r}_{r_p}\frac{1}{r'\sqrt{\left(1-\frac{r_{h}}{r'}\right)\left(r'-r_p\right)\left(r_a -r'\right)}}\,\,dr' - \frac{L_{\mathrm K}}{\sqrt{\mu^{2}-E^{2}}}\int^{r}_{r_p}\frac{1}{r'\sqrt{\left(r'-r_p\right)\left(r_a -r'\right)}}\,\,dr'\nonumber \\
 & =\frac{L}{\sqrt{\mu^{2}-E^{2}}}\sum^{\infty}_{n=0}\left(\begin{array}{c}
-\frac{1}{2}\\
n
\end{array}\right)\frac{r^{n}_{h}}{n!}\frac{\partial^{n}}{\partial r^{n}_{+}}\left[\frac{2\arctan\left(\sqrt{\frac{r_{+}+r_{-}}{r_{+}-r_{-}}}\sqrt{\frac{r-r_p}{r_a -r}}\right)}{\sqrt{r^{2}_{+}-r^{2}_{-}}}\right]
-\frac{2L_{{\rm K}}}{\sqrt{\mu^{2}-E^{2}}}\frac{\arctan\left(\sqrt{\frac{r_{a}}{r_{p}}}\sqrt{\frac{r-r_{p}}{r_{a}-r}}\right)}{\sqrt{r_{p}r_{a}}}.
\end{align}

Following the same approach used to determine $\Delta q^{r}$, 
$\left(\frac{\Omega^{\varphi}_{{\rm S}}}{\Omega^{r}_{{\rm S}}}-\frac{\Omega^{\varphi}_{{\rm K}}}{\Omega^{r}_{{\rm K}}}\right)m\,q^{r}_{{\rm K}}\left(r\right)-m\,\Delta\varphi\left(r\right)\biggr|_{r=r_{p},r_{a}}=0$ must hold,
resulting in a structure similar to that in~\eqref{appeq:deltaqr}, 
\begin{equation}
\left(\frac{\Omega^{\varphi}_{{\rm S}}}{\Omega^{r}_{{\rm S}}}-\frac{\Omega^{\varphi}_{{\rm K}}}{\Omega^{r}_{{\rm K}}}\right)m\,q^{r}_{{\rm K}}\left(r\right)-m\,\Delta\varphi\left(r\right)=m\,\Delta\chi\left(r\right)\sqrt{\left(r-r_{p}\right)\left(r_{a}-r\right)}-2\left(\frac{\Omega^{\varphi}_{{\rm S}}}{\Omega^{r}_{{\rm S}}}-\frac{\Omega^{\varphi}_{{\rm K}}}{\Omega^{r}_{{\rm K}}}\right)m\arctan\left[\frac{\sqrt{\left(r-r_{p}\right)\left(r_{a}-r\right)}}{r+\sqrt{r_{p}r_{a}}}\right]\,,
\end{equation}
where 
\begin{equation} \label{appeq:Deltachi}
\Delta\chi\left(r\right)=\frac{\left(e^{2}-1\right)(2e-p+2)(2e+p-2)}{p^{2}\left(e^{2}-p+3\right)}\left(\frac{\Omega^{\varphi}_{{\rm S}}}{\Omega^{r}_{{\rm S}}}-\frac{\Omega^{\varphi}_{{\rm K}}}{\Omega^{r}_{{\rm K}}}\right)-\sqrt{\frac{p}{p-4}}\frac{\sqrt{1-e^{2}}}{p}\sum^{\infty}_{k=1}\left(-1\right)^{k}\left(\begin{array}{c}
-\frac{1}{2}\\
k
\end{array}\right)\left(\frac{2\sqrt{1-e^{2}}}{p-4}\right)^{k}P^{\left(1\right)}_{k}\left(r\right)\,.
\end{equation}
Hence, we identify

\begin{equation} \label{appeq:chix2}
\chi^{\left(2\right)}_{1}\left(r\right)=\frac{\Omega^{\varphi}_{{\rm S}}}{\Omega^{r}_{{\rm S}}}\chi^{\left(1\right)}_{1}+\Delta\chi\left(r\right),\,\,\chi^{\left(2\right)}_{2}=\frac{\Omega^{\varphi}_{{\rm S}}}{\Omega^{r}_{{\rm S}}}\chi^{\left(1\right)}_{2}-2\left(\frac{\Omega^{\varphi}_{{\rm S}}}{\Omega^{r}_{{\rm S}}}-\frac{\Omega^{\varphi}_{{\rm K}}}{\Omega^{r}_{{\rm K}}}\right)\,\,.
\end{equation}

Finally, we note that the periastron advance in both Schwarzschild and Kepler are known explicitly,

\begin{equation}
   \frac{\Omega^{\varphi}_{{\rm K}}}{\Omega^{r}_{{\rm K}}} = \sqrt{\frac{4 e^2-(p-3) p}{4 e^2-(p-2)^2}},\quad \frac{\Omega^{\varphi}_{{\rm S}}}{\Omega^{r}_{{\rm S}}} = \frac{2}{\pi } \sqrt{\frac{p}{2 e+p-6}}\, K\left(\frac{4 e}{2 e+p-6}\right)\,,
\end{equation}
where $K\left(x\right)$ is the complete elliptic integral of the first kind.

\section{B.~$Z^{+}_{n \ell m}$: Definitions and technical details}
\subsection{B.1~ Definition of $\mathcal{D}_r$}
The flux Fourier coefficients $Z^{+}_{n \ell m}$ are well known in the literature in the form
\begin{align} \label{appeq:Z+o}
    Z^{+}_{n \ell m} &=\frac{1}{T_{\mathrm S}^r}\int_{0}^{T_{\mathrm S}^r}\,\overline{\mathcal{D}}_r R_{\ell m}^{\mathrm{in}}[r(t_{\mathrm S})]\,e^{i [\omega^{\mathrm S}_{nm}t_{\mathrm S}-m\varphi_{\mathrm S} (t_{\mathrm S})]}\,dt_{\mathrm S}\,.
\end{align}
Here $\overline{\mathcal{D}}_r$ is a second-order differential operator given explicitly in the next section. For now, we present it in the form
\begin{equation} \label{appeq:Dr12}
    \overline{\mathcal{D}}_r R_{\ell m}^{\mathrm{in}}= \mathcal{D}^{(1)}_r R_{\ell m}^{\mathrm{in}} + \frac{\sqrt{U^r_{\rm S}(r)}} {\mu} \mathcal{D}^{(2)}_r R_{\ell m}^{\mathrm{in}}\,.
\end{equation}
We wish to convert \eqref{appeq:Z+o} to a Fourier element of the form $\{\mathcal{D}_rR_{\ell m}^{\mathrm{in}}\}_S^{nm}$ where the curly brackets are defined in \eqref{SchFour} of the main text. To do so, we note that the $\sqrt{U^r_{\rm S}(r)}$ factor in \eqref{appeq:Dr12} changes sign half way through the period, namely when the radius is decreasing from $r_a$ to $r_p$. Consequently, the expression \eqref{appeq:Z+o} reduces to 
\begin{align} \label{appeq:Z+}
    Z^{+}_{n \ell m} &= \left\{\mathcal{D}^{(1)}_r R_{\ell m}^{\mathrm{in}}\right\}^{nm}_{\rm S} + \frac{i}{\pi \mu}\int_{r_{p}}^{r_{a}}\sqrt{U^r_{\rm S}(r)}\, \mathcal{D}^{(2)}_r R_{\ell m}^{\mathrm{in}}\,\sin[ A^{\rm S}_{nm} (r)]\,t'_{\rm S}(r)\,dr,\\
    &= \left\{\mathcal{D}^{(1)}_r R_{\ell m}^{\mathrm{in}}\right\}^{nm}_{\rm S} + \frac{i}{\pi}\int_{r_{p}}^{r_{a}}\frac{\mathcal{E}r^{4}}{\Delta(r)}\, \mathcal{D}^{(2)}_r R_{\ell m}^{\mathrm{in}}\,\sin[ A^{\rm S}_{nm} (r)]\,dr,
\end{align}
where in the second line we used~\eqref{EOMSch} of the main text. In order to deal with the $\sin$ term, we convert it to a $\cos$ term using integration-by-parts, resulting in
\begin{align} \label{appeq: Z+f}
    Z^{+}_{n \ell m} &= \left\{\mathcal{D}^{(1)}_r R_{\ell m}^{\mathrm{in}}\right\}^{nm}_{\rm S} - \frac{i}{\pi}\int_{r_{p}}^{r_{a}} \mathcal{F}(r) \left(\omega^{\rm S}_{nm} - m\,\frac{\mathcal{L} \Delta(r)}{\mathcal{E} r^4}\right) \,\cos[ A^{\rm S}_{nm} (r)]\,t'_{\rm S}(r)\,dr\nonumber\\
    &= \left\{\mathcal{D}_r R_{\ell m}^{\mathrm{in}}\right\}^{nm}_{\rm S},
\end{align}
where
\begin{align} \label{appeq: DDr}
    \mathcal{D}_r R_{\ell m}^{\mathrm{in}} &= \mathcal{D}^{(1)}_r R_{\ell m}^{\mathrm{in}} - i  \left(\omega^{\rm S}_{nm} - m\,\frac{\mathcal{L} \Delta(r)}{\mathcal{E} r^4}\right)\int_{r_p}^{r} \frac{\mathcal{E}r'^{4}}{\Delta(r')}\mathcal{D}^{(2)}_r R_{\ell m}^{\mathrm{in}}\left(r'\right)\,dr'\,.
\end{align}
\subsection{B.2~Explicit Expressions for $\mathcal{D}^{(1,2)}_r$}
In this subsection, we state the definition of the linear differential operator $\overline{\mathcal{D}}_r$. In the planar Schwarzschild case that we consider in this work, $\overline{\mathcal{D}}_r$ is given by
\begin{align} \label{appeq:Dr}
    \overline{\mathcal{D}}_r = \frac{4\pi}{W_r} \sum_{i,j=1}^{3} \Xi_{ij}\left(r\right) \frac{\partial^{j-1}}{\partial\theta^{j-1}}\Big[{}_{-2}Y_{\ell m}\left(\theta,0\right)\Big] \Bigg|_{\theta=0} \frac{d^{i-1}}{dr^{i-1}},
\end{align}
where $W_r$ is the ($r-$independent) Wronksian associated with the upgoing and ingoing radial solutions of the homogeneous Teukolsky equation,
\begin{equation}
    W_r\equiv \Delta\left(r\right)^{-1}\left(R_{\ell m}^{\mathrm{in}} \frac{d}{dr}R_{\ell m}^{\mathrm{up}} - R_{\ell m}^{\mathrm{up}} \frac{d}{dr}R_{\ell m}^{\mathrm{in}} \right),
\end{equation}
and $\Xi_{ij}$ are defined as
\begin{subequations} \label{appeq:Zij}
    \begin{align}
      \Xi_{11}\equiv &
\frac{1}{4(r-2)r^5\mathcal{E}}
\Bigg\{2\sqrt{r^4\mathcal{E}^2-(r-2)r\left(r^2+\mathcal{L}^2\right)}
\left[r^2\left(m\omega\mathcal{L}-(m^2-2)\mathcal{E}\right)
-2im(r-2)\mathcal{L}
\right] \nonumber\\
&\qquad +r\Big[r^3\left(-(m^2-2)(2\mathcal{E}^2-1)
+2m\omega\mathcal{E}\mathcal{L}
-\omega^2\mathcal{L}^2\right)-2r^2\left(
m^2+2im\mathcal{E}\mathcal{L}
-i\omega\mathcal{L}^2-2
\right) \nonumber\\
&\qquad\qquad
+r\mathcal{L}\left(m^2\mathcal{L}+8im\mathcal{E}-2i\omega\mathcal{L}-2\mathcal{L}
\right)-2(m^2-2)\mathcal{L}^2
\Big]\Bigg\},\\
\Xi_{12}\equiv & \frac{1}{2(r-2)r^5\mathcal{E}}
\Bigg\{\sqrt{r^4\mathcal{E}^2-(r-2)r\left(r^2+\mathcal{L}^2\right)}
\left[2mr^2\mathcal{E} -r^2\omega\mathcal{L}
+2i(r-2)\mathcal{L}\right] \nonumber\\
&\qquad-r\Big[
mr^2\left(-2r\mathcal{E}^2+r-2\right)
+m(r-2)\mathcal{L}^2 +r\mathcal{E}\mathcal{L}
\left(r(r\omega-2i)+4i\right)\Big]\Bigg\},\\
     \Xi_{13}\equiv & \frac{-2 r^3 \mathcal{E}^2+r^3-2 r^2-2 r \mathcal{E} \sqrt{r^4
   \mathcal{E}^2-(r-2) r \left(r^2+\mathcal{L}^2\right)}+r \mathcal{L}^2-2
   \mathcal{L}^2}{4 (r-2) r^4 \mathcal{E}},\\
   \Xi_{21}\equiv & \frac{\mathcal{L} \left(i r^2 (m \mathcal{E}-\omega  \mathcal{L})+i m
   \sqrt{r^4 \mathcal{E}^2-(r-2) r \left(r^2+\mathcal{L}^2\right)}-r
   \mathcal{L}+2 \mathcal{L}\right)}{2 r^4 \mathcal{E}},\\
   \Xi_{22}\equiv & -\frac{i \mathcal{L} \left(r^2 \mathcal{E}+\sqrt{r^4 \mathcal{E}^2-(r-2) r
   \left(r^2+\mathcal{L}^2\right)}\right)}{2 r^4 \mathcal{E}},\\
   \Xi_{31}\equiv & \frac{(r-2) \mathcal{L}^2}{4 r^3 \mathcal{E}},\qquad
   \Xi_{23} = \Xi_{32} = \Xi_{33} = 0,
    \end{align}
\end{subequations}
where $\omega=\omega_{nm}^{\rm S}$, $\mathcal{E} = E/\mu$ and $\mathcal{L} = L_S/\mu$. Extracting the coefficients multiplying the square root in the obvious way, we get the operators $\mathcal{D}^{(1)}_r$ and $\mathcal{D}^{(2)}_r$ relevant for \eqref{appeq: DDr}.

\section{C.~Toy calculation}
In this section, we explicitly demonstrate the calculations done in this work, taking the various cutoffs to be small for compactness of the resulting expressions. This should be regarded as a toy calculation, intended to highlight the key aspects of the algorithm proposed here. Scaling this calculation yields the benchmarks reported in Fig.~\ref{fig:fluxes} of the main text. 
\subsection{C.1~Laurent-like expansion of $\mathcal{K}[f]$}
In this calculation, we truncate, when necessary, the PN-compatible pieces such as $\chi^{\left(0\right)}\left(r\right)$ and the $\chi_1$'s at $2$PN order. Starting with $\chi^{\left(0\right)}$, this corresponds to truncating~\eqref{appeq:qsandchi0} at $\mathcal{O}\left(r^{-3}\right)$,
\begin{align} \label{appeq: chi0exp}
&\chi^{\left(0\right)}\left(r\right) = \beta_0 + \frac{\beta_1}{r} + \frac{\beta_2}{r^2} +\mathcal{O}\left(r^{-3}\right),\nonumber\\
\beta_0 &= \frac{E}{\sqrt{\mu^2-E^2}},\quad \frac{\beta_1}{\beta_0} = 3+\frac{4}{p-4},\quad  \frac{\beta_2}{\beta_0} = \frac{5 p (3 p-16)+128}{2 (p-4)^2}.
\end{align}
Next, we turn to $\chi^{\left(1\right)}_{1}$. At this order, this corresponds to cutting out the sum in~\eqref{appeq: chi11},
\begin{equation}\label{toy:chi11at2PN}
\chi^{\left(1\right)}_{1}=\frac{\left(1-e^{2}\right)\left(2+2e-p\right)\left(2e+p-2\right)}{p^{2}\left(3+e^{2}-p\right)}-\Omega^{r}_{{\rm Sch}}\beta_{0} + \mathcal{O}\left(p^{-3}\right)\,.
\end{equation}
Note that at this order, $\chi_1^{(1)}$ is $r-$independent. Truncating the sum in~\eqref{appeq:Deltachi} at $k=2$ and using~\eqref{appeq:chix2} yields

\begin{align} \label{toy:chi12at2PN}   \chi_1^{(2)}\left(r\right) =& -\frac{3 \sqrt{1-e^2} p^{3/2} \left(\frac{1}{p-4}\right)^{5/2}}{4
   r^2}-\frac{\left(1-e^2\right) \sqrt{\left((p-2)^2-4 e^2\right)
   \left((p-3) p-4 e^2\right)}}{p^2 \left(e^2-p+3\right)}\nonumber\\
   &+\frac{\sqrt{1-e^2}
   (7-4 p) \sqrt{p} \left(\frac{1}{p-4}\right)^{5/2}}{4 r} - \Omega^{\varphi}_{{\rm Sch}}\beta_{0} + \mathcal{O}\left(r^{-3}\right).
\end{align}
Using~\eqref{appeq:deltaqr} and~\eqref{appeq:chix2} for $\chi_2^{(1,2)}$ along with~\eqref{toy:chi11at2PN} and~\eqref{toy:chi12at2PN} fully determines $\Delta A_{nm}$ at this order,
\begin{eqnarray}
    &&\Delta A_{n m}(r) ={\sqrt{\left(r-r_{p}\right)\left(r_{a}-r\right)}} \left[ \chi_1^{(1)}\left(r\right) n + \chi_1^{(2)}\left(r\right) m + \chi_2^{nm}\,g(r)\right],
\end{eqnarray}
where $g(r)$ is defined in~\eqref{eqg} of the main text, and we defined $\chi_2^{nm} \equiv \left(\chi_2^{(1)} n + \chi_2^{(2)} m\right)$.\\
\indent We now turn to finding a Laurent-like expansion of $\mathcal{A}_{nm}$. To this, we employ the Chebyshev expansion~\eqref{eq:cossinCheb} of $\cos$ with $N_{\cos}=1$, 
\begin{equation} \label{appeq:coscheb}
    \cos\left(x\right) = J_0(a)+2 J_2(a) -\frac{4 x^2}{a^2}  J_2(a)+\dots,
\end{equation}
where $a$ is an arbitrary parameter that we will choose shortly, and ``$\dots$'' denotes dropped terms. Coupled with~\eqref{appeq: chi0exp}, we obtain the Laurent-like expansion of $\mathcal{A}_{nm}$,
\begin{subequations} \label{appeq:AnmLL}
\begin{align}
    \mathcal{A}_{nm}\left(r\right) = \beta_0^{-1}\chi^{(0)} \left(r\right) \cos\left(\Delta A_{nm}\right) ={}&
\mathcal E_3r^3
+\mathcal E_2r^2
+\mathcal E_1r
+\mathcal E_0
+\frac{\mathcal E_{-1}}{r}
+\frac{\mathcal E_{-2}}{r^2} +\dots,
\end{align}
\begin{align}
\mathcal E_3
={}&
\frac{4J_2(a)}{a^2}\tilde{\mathcal{E}}_1, \qquad
\mathcal E_2
=
\frac{4J_2(a)}{a^2}
\left[
\tilde{\mathcal{E}}_0-\frac{2p}{q_e^2}\tilde{\mathcal{E}}_1
\right],
\\
\mathcal E_1
={}&
\frac{4J_2(a)}{a^2}
\left[
\tilde{\mathcal{E}}_{-1}
-\frac{2p}{q_e^2}\tilde{\mathcal{E}}_0
+\frac{p^2}{q_e^2}\tilde{\mathcal{E}}_1
\right],
\\
\mathcal E_0
={}&
J_0(a)+2J_2(a)
+
\frac{4J_2(a)}{a^2}
\left[
\tilde{\mathcal{E}}_{-2}
-\frac{2p}{q_e^2}\tilde{\mathcal{E}}_{-1}
+\frac{p^2}{q_e^2}\tilde{\mathcal{E}}_0
\right],
\\
\mathcal E_{-1}
={}&
\frac{\beta_1}{\beta_0}
\left[J_0(a)+2J_2(a)\right]
+
\frac{4J_2(a)}{a^2}
\left[
-\frac{2p}{q_e^2}\tilde{\mathcal{E}}_{-2}
+\frac{p^2}{q_e^2}\tilde{\mathcal{E}}_{-1}
\right],
\\
\mathcal E_{-2}
={}&
\frac{\beta_2}{\beta_0}
\left[J_0(a)+2J_2(a)\right]
+
\frac{4J_2(a)}{a^2}
\frac{p^2}{q_e^2} 
\tilde{\mathcal{E}}_{-2}.
\end{align}
\end{subequations}
Here we identify several building blocks, given by
\allowdisplaybreaks
\begin{subequations}
\begin{align}
\tilde{\mathcal{E}}_1
={}&
-\frac{2 \left[\chi_2^{nm}\right]^2 q_e^6}{(1+q_e)^3p^3}
+
\frac{2 \chi_2^{nm} \beta_0 \left(\omega^{\rm S}_{nm}-\omega^{\rm K}_{nm}\right) q_e^4}{(1+q_e)^2p^2},
\\[1ex]
\tilde{\mathcal{E}}_0
={}&
\beta_0^2 \left(\omega^{\rm S}_{nm}-\omega^{\rm K}_{nm}\right)^2
+
\left[\chi_2^{nm}\right]^2
\left[
\frac{q_e^4(3q_e+13)}{4(1+q_e)^3p^2}
-
\frac{2\beta_1q_e^6}{\beta_0(1+q_e)^3p^3}
\right]
\nonumber\\
&-
2\chi_2^{nm}
\left[
\frac{\beta_0 \left(\omega^{\rm S}_{nm}-\omega^{\rm K}_{nm}\right) q_e^2 (q_e+2)}{(1+q_e)^2p}
-
\frac{
\left(
 \beta_1 \left(\omega^{\rm S}_{nm}-\omega^{\rm K}_{nm}\right)
+
\frac{m q_e\sqrt p\,(4p-7)}{4(p-4)^{5/2}}
\right)q_e^4
}{
(1+q_e)^2p^2
}
\right],
\\[1ex]
\tilde{\mathcal{E}}_{-1}
=&
 \beta_1 \beta_0 \left(\omega^{\rm S}_{nm}-\omega^{\rm K}_{nm}\right)^2
+
\frac{mq_e\sqrt p\,(4p-7)}{2(p-4)^{5/2}} \beta_0 \left(\omega^{\rm S}_{nm}-\omega^{\rm K}_{nm}\right) +
\left[\chi_2^{nm}\right]^2
\left[
\frac{\beta_1q_e^4(3q_e+13)}
{4\beta_0(1+q_e)^3p^2}
-
\frac{2\beta_2q_e^6}
{\beta_0(1+q_e)^3p^3}
\right]
\nonumber\\
&-
2\chi_2^{nm}
\left[
\frac{
\left(
 \beta_1 \left(\omega^{\rm S}_{nm}-\omega^{\rm K}_{nm}\right)
+
\frac{mq_e\sqrt p\,(4p-7)}{4(p-4)^{5/2}}
\right)q_e^2(q_e+2)
}{
(1+q_e)^2p
}
-
\frac{
\left(
 \beta_2 \left(\omega^{\rm S}_{nm}-\omega^{\rm K}_{nm}\right)
+
\frac{mq_e\sqrt p}{4(p-4)^{5/2}}
\left[
\frac{(4p-7)\beta_1}{\beta_0}+3p
\right]
\right)q_e^4
}{
(1+q_e)^2p^2
}
\right],
\\[1ex]
\tilde{\mathcal{E}}_{-2}
={}&
\beta_2\beta_0 \left(\omega^{\rm S}_{nm}-\omega^{\rm K}_{nm}\right)^2
+
\frac{mq_e\sqrt p}{2(p-4)^{5/2}}
\big[
(4p-7)\beta_1 + 3p \beta_0
\big] \left(\omega^{\rm S}_{nm}-\omega^{\rm K}_{nm}\right) +
\frac{m^2q_e^2p(4p-7)^2}{16(p-4)^5}
\nonumber\\
&
+
\frac{\beta_2\left[\chi_2^{nm}\right]^2q_e^4(3q_e+13)}
{4\beta_0(1+q_e)^3p^2} -
\frac{
2\chi_2^{nm}
\left(
\beta_2 \left(\omega^{\rm S}_{nm}-\omega^{\rm K}_{nm}\right)
+
\frac{m q_e\sqrt p}{4(p-4)^{5/2}}
\left[
\frac{(4p-7)\beta_1}{\beta_0}+3p
\right]
\right)q_e^2(q_e+2)
}{
(1+q_e)^2p
},
\end{align}
\end{subequations}
where $q_e=\sqrt{1-e^2}$. In order to obtain~\eqref{appeq:AnmLL}, we also PN-expanded the higher powers of $\chi_1^{(1,2)}$ encountered when using~\eqref{appeq:coscheb}, up to $2$PN. We also used the Chebyshev expansion of $g(r)$ and $g^2(r)$. For $g(r)$ we truncated Eq.~\eqref{gcheb} at $i=1$,
\begin{equation} \label{appeq:gcheb}
    g\left(r\right) = \frac{2 (1-q_e)}{\left(r_{a}-r_{p}\right)e} \left(1-\frac{1-q_e}{2e} U_1\left(\frac{2 r-r_{a}-r_{p}}{r_{a}-r_{p}}\right)\right)+\dots = \frac{q_e^2 (q_e+2)}{p (q_e+1)^2}-\frac{q_e^4 r}{p^2 (q_e+1)^2} + \dots
\end{equation}
The expansion for $g^2(r)$ is obtained from~\eqref{appeq:gcheb} using the identity $U_i(x) U_j(x)=\sum_{s=0}^{\min (i, j)} U_{i+j-2 s}(x)$, keeping only the terms up to $U_1$ for simplicity. This yields
\begin{equation} \label{appeq:gcheb2}
    g^2\left(r\right) = \frac{q_e^4 \left(3 q_e+13\right)}{4 p^2 \left(q_e+1\right){}^3}-\frac{2 r q_e^6}{p^3 \left(q_e+1\right){}^3} + \dots
\end{equation}
\indent We now move on to $\mathcal{C}_{nm}$. Finding the Laurent-like expansion of it is similar to $\mathcal{A}_{nm}$, except we use the Chebyshev expansion of $\sin$ in~\eqref{eq:cossinCheb}, truncated at $i=0$
\begin{equation} \label{appeq:sincheb}
    \sin\left(x\right) = \frac{2 J_1\left(a\right)}{a} x +\dots
\end{equation}
The resulting Laurent-like expansion is given by
\allowdisplaybreaks
\begin{subequations} \label{appeq:Cnm}
\begin{align}
\mathcal{C}_{nm}(r) ={}& \beta_0^{-1} \chi^{(0)}(r)\frac{r\sin\left(\Delta A_{nm}(r)\right)}{\sqrt{\left(r-r_{p}\right)\left(r_{a}-r\right)}}
=
\tilde{\mathcal{F}}_2 r^2
+\tilde{\mathcal{F}}_1 r
+\tilde{\mathcal{F}}_0
+\frac{\tilde{\mathcal{F}}_{-1}}{r} + \dots,
\\[1ex]
\tilde{\mathcal{F}}_2
={}&
-\frac{2J_1(a)}{a}
\frac{q^4 \chi_2^{nm}}{p^2(q+1)^2},
\\[1ex]
\tilde{\mathcal{F}}_1
={}&
\frac{2J_1(a)}{a}
\left[
\frac{q^2(q+2) \chi_2^{nm}}{p(q+1)^2}
-\beta_0 \left(\omega^{\rm S}_{nm}-\omega^{\rm K}_{nm}\right)
-\frac{\beta_1}{\beta_0}
\frac{q^4 \chi_2^{nm}}{p^2(q+1)^2}
\right],
\\[1ex]
\tilde{\mathcal{F}}_0
={}&
\frac{2J_1(a)}{a}
\left[
\frac{\beta_1}{\beta_0}
\left(
\frac{q^2(q+2) \chi_2^{nm}}{p(q+1)^2}
-\beta_0 \left(\omega^{\rm S}_{nm}-\omega^{\rm K}_{nm}\right)
\right)
-\frac{\beta_2}{\beta_0}
\frac{q^4 \chi_2^{nm}}{p^2(q+1)^2}
-\frac{m q\sqrt p\,(4p-7)}{4(p-4)^{5/2}}
\right],
\\[1ex]
\tilde{\mathcal{F}}_{-1}
={}&
\frac{2J_1(a)}{a}
\left[
\frac{\beta_2}{\beta_0}
\left(
\frac{q^2(q+2) \chi_2^{nm}}{p(q+1)^2}
-\beta_0 \left(\omega^{\rm S}_{nm}-\omega^{\rm K}_{nm}\right)
\right)
-\frac{mq\sqrt p}{4(p-4)^{5/2}}
\left(
\frac{\beta_1}{\beta_0}(4p-7)+3p
\right)
\right].
\end{align}
\end{subequations}
\subsection{C.3~ Computation of $Z^+_{n,\ell=2,m=2}$}
We begin with the $2$PN Teukolsky radial modes obtained through BHPT~\cite{BHPToolkit, Castillo:2024isq}, rewritten as a Laurent series in $r$,
\begin{subequations} \label{appeq: Teukrad}
\begin{align}
  R_{\ell=2,m=2}^{\mathrm{in}} \left(r\right) ={}& \frac{16\omega^4}{5} r^2
-\frac{16\omega^4}{5} r^3
+\frac{2\omega^4}{5}\left(2-i \left(-3+4 \gamma \right) \omega\right) r^4
+\frac{8 \omega^5}{315} \left(21 i+\left(-73+42 \gamma \right) \omega\right) r^5\nonumber\\
&\qquad \qquad -\frac{22 \omega^6}{105} r^6
-\frac{2i \omega^7}{35} r^7
+\frac{23\omega^8}{1890} r^8+\dots,
\end{align}
\begin{align}
  R_{\ell=2,m=2}^{\mathrm{up}} \left(r\right) ={}& -\frac{3 i}{r^2 \omega }+\frac{-\frac{3 i}{\omega }+2-6 \gamma +6 i \pi }{r}-3 +\frac{3 i r \omega }{2} + \frac{r^2 \omega ^2}{2} +\dots,
\end{align}
\end{subequations}
where $\gamma$ is the Euler-Mascheroni constant, and $\omega = \omega_{n2}^{\mathrm S}$.
The corresponding Wronksian is given by
\begin{equation}
    W_r = 6 (11-8 \gamma +4 i \pi ) \omega ^4 - 12 i \omega ^3 + \dots
\end{equation}
The radial modes~\eqref{appeq: Teukrad} have a different normalization condition than that required for~\eqref{appeq:Dr12} and~\eqref{appeq:Z+}. To compensate, we multiply by the relevant transmission amplitude  $C_{\rm trans}$ (also computed using the BHPT) given by
\begin{equation}
    C_{\rm trans} = \frac{2}{3} \omega ^3 (\omega  (6 \log (\omega )-1+3 i \pi +\log (4096))-3 i)+\dots
\end{equation}
We absorb this coefficient in $\mathcal{D}_r^{(1)}$ and $\mathcal{D}_r^{(2)}$. Expanding the $\Xi_{ij}$ coefficients~\eqref{appeq:Zij} to next-to-next-to-leading-order (NNLO) in $1/r$, we get
\begin{subequations} \label{appeq:Dr1R}
\begin{align}
\mathcal{D}^{(1)}_r R_{{\ell=2,m=2}}^{\mathrm{in}}
={}&
\frac{\pi C_{\rm trans}\,\omega^4}{3024\sqrt{5\pi}\,W_r \mathcal E}
\left(
\sigma^{(1)}_6 r^6
+\sigma^{(1)}_5 r^5
+\sigma^{(1)}_4 r^4
+\sigma^{(1)}_3 r^3
+\sigma^{(1)}_2 r^2
+\sigma^{(1)}_1 r
+\sigma^{(1)}_0
+\frac{\sigma^{(1)}_{-1}}{r}
+\frac{\sigma^{(1)}_{-2}}{r^2}
+\frac{\sigma^{(1)}_{-3}}{r^3}+\dots
\right),
\\[1ex]
\sigma^{(1)}_6
={}&
-23\omega^4
\left(
24\mathcal E^2
-8\omega\mathcal E\mathcal L
+\omega^2\mathcal L^2
-12
\right),
\\
\sigma^{(1)}_5
={}&
-2\omega^3
\left(
24(23\omega-54i)\mathcal E^2
-8\omega(23\omega+15i)\mathcal E\mathcal L
+\omega^2(23\omega+107i)\mathcal L^2
+648i
\right),
\\
\sigma^{(1)}_4
={}&
2\omega^2
\left(
-48(23\omega^2-54i\omega-99)\mathcal E^2
+16\omega(23\omega^2-54i\omega+36)\mathcal E\mathcal L
\right.
\nonumber\\
&\hspace{3.2cm}
\left.
+\omega^2(-46\omega^2+131i\omega+148)\mathcal L^2
-2376
\right),
\\
\sigma^{(1)}_3
={}&
8\omega
\left(
72(42\gamma-73)\omega
+72\left(18i\omega^2+(179-84\gamma)\omega-42i\right)\mathcal E^2
\right.
\nonumber\\
&\hspace{2.4cm}
+24\left(-18i\omega^2+(84\gamma-179)\omega-24i\right)
\omega\mathcal E\mathcal L
\nonumber\\
&\hspace{2.4cm}
\left.
+\left(54i\omega^4+2(167-126\gamma)\omega^3-171i\omega^2\right)\mathcal L^2
+1512i
\right),
\\
\sigma^{(1)}_2
={}&
-12
\left(
756(4\gamma-3)i\omega
+24\left(4(84\gamma-179)\omega^2-21i(12\gamma-17)\omega+126\right)\mathcal E^2
\right.
\nonumber\\
&\hspace{2.2cm}
-8\left(4(84\gamma-179)\omega^2+3(84\gamma-173)i\omega-126\right)
\omega\mathcal E\mathcal L
\nonumber\\
&\hspace{2.2cm}
\left.
+\left(4(84\gamma-179)\omega^4+(1092\gamma-2417)i\omega^3
+444\omega^2\right)\mathcal L^2
-1512
\right),
\\
\sigma^{(1)}_1
={}&
24
\left(
-24\left(8(42\gamma-73)\omega^2-21i(12\gamma-17)\omega-126\right)\mathcal E^2
\right.
\nonumber\\
&\hspace{2.1cm}
+8\left(
8(42\gamma-73)\omega^3
-21i(12\gamma-17)\omega^2
+63(4\gamma-5)\omega
+126i
\right)\mathcal E\mathcal L
\nonumber\\
&\hspace{2.1cm}
\left.
+\omega\left(
-8(42\gamma-73)\omega^3
+(420\gamma-649)i\omega^2
+7(156\gamma-287)\omega
+546i
\right)\mathcal L^2
-3024
\right),
\\
\sigma^{(1)}_0
={}&
24i
\left(
3024(4\gamma-3)\omega\mathcal E^2
-1008\left((4\gamma-3)\omega^2+2\right)\mathcal E\mathcal L
\right.
\nonumber\\
&\hspace{2.0cm}
\left.
+\left(
126(4\gamma-3)\omega^3
+(1428\gamma-2731)i\omega^2
-42(48\gamma-43)\omega
-1008i
\right)\mathcal L^2
-3024i
\right),
\\
\sigma^{(1)}_{-1}
={}&
6048
\left(
48\mathcal E^2
-16\omega\mathcal E\mathcal L
+\left(2\omega^2+(4\gamma-7)i\omega-14\right)\mathcal L^2
\right),
\\
\sigma^{(1)}_{-2}
={}&
12096
\left(
-48\mathcal E^2
+16\omega\mathcal E\mathcal L
+\left(-2\omega^2+i\omega+5\right)\mathcal L^2
\right),
\\
\sigma^{(1)}_{-3}
={}&
24192\mathcal L^2 .
\end{align}
\end{subequations}
Similarly, one obtains an expression for the integrand in~\eqref{appeq: DDr} up to the same order. Integrating this expression $\mathcal{F}$ becomes trivial, and the constant term can be dropped since it does not contribute. Upon integration, the $\frac{1}{r}$ term results in a $\log{r}$ term. To obtain a Laurent-like form, we replace it with the following truncated Gegenbauer expansion ($k=1$ in~\eqref{eq:gegen} of the main text),
\begin{equation}
    \log{r} = \frac{2 q_e^2}{p \left(q_e+1\right)} r +\log \left(\frac{p \left(q_e+1\right)}{2
   q_e^2}\right)-\frac{2}{q_e+1} +\dots
\end{equation}
For notational simplicity, denote $\ell_e
=\log\!\left(\frac{p(q_e+1)}{2q_e^2}\right)$ and $\kappa_e = 12\mathcal E+\mathcal L\bigl((4\gamma-7)\omega+i\bigr)$. The resulting Laurent-like form is given by
\allowdisplaybreaks
\begin{subequations}\label{appeq:curlyF-laurent}
\begin{align}
\mathcal{F}(r) \left(\omega - \,\frac{2\mathcal{L} \Delta(r)}{\mathcal{E} r^4}\right)
={}&
\frac{4\pi i C_{\rm trans}\,\omega^4}{\sqrt{5\pi}\,W_r \mathcal E}
\biggl(
\sigma^{(2)}_7 r^7
+\sigma^{(2)}_6 r^6
+\sigma^{(2)}_5 r^5
+\sigma^{(2)}_4 r^4
+\sigma^{(2)}_3 r^3
+\sigma^{(2)}_2 r^2
+\sigma^{(2)}_1 r
+\sigma^{(2)}_0
\nonumber\\
&\quad
+\frac{\sigma^{(2)}_{-1}}{r}
+\frac{\sigma^{(2)}_{-2}}{r^2}
+\frac{\sigma^{(2)}_{-3}}{r^3}
+\frac{\sigma^{(2)}_{-4}}{r^4}
+\frac{\sigma^{(2)}_{-5}}{r^5} + \dots
\biggr).
\end{align}
\begin{align}
\sigma^{(2)}_7
={}&
\frac{23i\,\mathcal E\,\omega^5}{10584}
\bigl(3\mathcal E-\mathcal L\omega\bigr),
\\
\sigma^{(2)}_6
={}&
\frac{\mathcal E\omega^4}{4536}
\bigl[
6\mathcal E(23i\omega+27)
+\mathcal L(15-46i\omega)\omega
\bigr],
\\
\sigma^{(2)}_5
={}&
\frac{i\omega^3}{26460}
\bigl[
126(23\omega^2-36i\omega-33)\mathcal E^2
-3\mathcal L\omega(322\omega^2-182i\omega+283)\mathcal E
+115\mathcal L^2\omega^2
\bigr],
\\
\sigma^{(2)}_4
={}&
-\frac{\omega^2}{15876}
\biggl[
-378\bigl(27\omega^2+2i(21\gamma-53)\omega-21\bigr)\mathcal E^2
\nonumber\\
&\quad
+3\mathcal L\omega
\bigl(490\omega^2+i(1764\gamma-3323)\omega+882\bigr)\mathcal E
+\mathcal L^2(105-184i\omega)\omega^2
\biggr],\\
% \end{align}
% \begin{align}
\sigma^{(2)}_3
={}&
\frac{i\omega}{5670}
\biggl[
45\bigl(4(168\gamma-391)\omega^2-21i(12\gamma-25)\omega+126\bigr)
\mathcal E^2
\nonumber\\
&\quad
-3\mathcal L\omega
\bigl(8(420\gamma-797)\omega^2
+3i(420\gamma-1151)\omega-1224\bigr)\mathcal E
\nonumber\\
&\quad
+\mathcal L^2\omega^2(184\omega^2-309i\omega+216)
\biggr],
\\
\sigma^{(2)}_2
={}&
\frac{\omega}{1890}
\biggl[
270\bigl(84\gamma(2i\omega+1)-292i\omega-147\bigr)\omega\mathcal E^2
\nonumber\\
&\quad
+9\mathcal L
\bigl(-12i(140\gamma-251)\omega^3-776\omega^2
-i(840\gamma-1663)\omega+420\bigr)\mathcal E
\nonumber\\
&\quad
+\mathcal L^2\omega
\bigl(-276i\omega^3+194\omega^2+3i(420\gamma-883)\omega+360\bigr)
\biggr],\\
\sigma^{(2)}_1
={}&
-\frac{1}{189}
\biggl[
\mathcal L^2\omega
\bigl(70\omega^3-i(420\gamma-811)\omega^2
+3(84\gamma-181)\omega+126i\bigr)
\nonumber\\
&\quad
+3\mathcal E\mathcal L
\bigl(-56(3\gamma-29)\omega^3
+2i(462\gamma-1073)\omega^2
+21(12\gamma-19)\omega+126i\bigr)
\nonumber\\
&\quad
-3402(4\gamma-3)\mathcal E^2\omega^2
\biggr]
-\frac{16i\,\mathcal E\omega\,\kappa_e q_e^2}{p(q_e+1)},
\\
\sigma^{(2)}_0
={}&
\frac{1}{189(q_e+1)}
\biggl[
i\biggl(
36288\omega\mathcal E^2
+84\mathcal L
\bigl(4(21\gamma-44)\omega^2+3i(12\gamma-7)\omega+9\bigr)\mathcal E
\nonumber\\
&\quad
+\mathcal L^2
\bigl(56(30\gamma-47)\omega^3
-12i(42\gamma-5)\omega^2
+63(12\gamma-17)\omega+378i\bigr)
\biggr)
\nonumber\\
&\quad
+\mathcal L q_e
\biggl(
84\mathcal E
\bigl(-4i(15\gamma-19)\omega^2+(57-36\gamma)\omega+9i\bigr)
\nonumber\\
&\quad
+\mathcal L
\bigl(56i(30\gamma-47)\omega^3
+12(42\gamma-5)\omega^2
+63i(12\gamma-17)\omega-378\bigr)
\biggr)
\nonumber\\
&\quad
-1512i\,\mathcal E\omega\,\kappa_e\ell_e(q_e+1)
\biggr],\\
\sigma^{(2)}_{-1}
={}&
-\frac{2i}{63}
\biggl[
4536\omega\mathcal E^2
-36\mathcal L\omega
\bigl(-104\omega+84\gamma(\omega+i)-63i\bigr)\mathcal E
\nonumber\\
&\quad
+\mathcal L^2
\bigl(24(42\gamma-73)\omega^3
-4i(42\gamma-349)\omega^2
+(777-252\gamma)\omega+126i\bigr)
\biggr]
+\frac{32i\,\mathcal L\kappa_e q_e^2}{p(q_e+1)},
\\
\sigma^{(2)}_{-2}
={}&
\frac{8\mathcal L}{63p(q_e+1)}
\biggl[
-504i\,\kappa_e q_e^2
+p\omega
\bigl(63(36\gamma-29)\mathcal E
+\mathcal L(-893\omega+84\gamma(\omega-3i)+609i)\bigr)q_e
\nonumber\\
&\quad
+p\bigl(
63\mathcal E((36\gamma-29)\omega-48i)
+\mathcal L((84\gamma-893)\omega^2
-21i(60\gamma-113)\omega+252)
\bigr)
\nonumber\\
&\quad
+126ip\,\kappa_e\ell_e(q_e+1)
\biggr],
\\
\sigma^{(2)}_{-3}
={}&
\frac{32\mathcal L}{q_e+1}
\biggl[
i\bigl(33\mathcal E+\mathcal L((8\gamma-17)\omega+5i)\bigr)
+(9i\mathcal E-3\mathcal L-3i\mathcal L\omega)q_e
\nonumber\\
&\quad
-i\kappa_e\ell_e(q_e+1)
\biggr],
\\
\sigma^{(2)}_{-4}
={}&
32\mathcal L
\bigl[\mathcal L(6i\omega+7)-18i\mathcal E\bigr],
\\
\sigma^{(2)}_{-5}
={}&
-64\mathcal L^2 .
\end{align}
\end{subequations}
Finally, assembling the results from~\eqref{appeq:Dr1R} and~\eqref{appeq:curlyF-laurent} yields the $f(r)$ needed to calculate $Z_{n22}$ according to~\eqref{appeq: Z+f}. Define $\Sigma_j = \sigma_j^{(1)}+12096\sigma_j^{(2)}$, with the convention that if a $\sigma_j$ does not explicitly appear in the expressions above, then it is taken as $0$. Using the explicit results in~\eqref{appeq:AnmLL} and~\eqref{appeq:Cnm}, we obtain at NNLO

\allowdisplaybreaks

\begin{subequations}\label{eq:K-laurent}
\begin{align}
\mathcal{K}\left[f\right]
={}&
\frac{C_{\rm trans}\sqrt{\pi/5}\,\omega^4}
{7620480\,W_r\,E}
\biggl(
\mathcal K_{10}r^{10}
+\mathcal K_9r^9
+\mathcal K_8r^8
+\mathcal K_7r^7
+\mathcal K_6r^6
+\mathcal K_5r^5
+\mathcal K_4r^4
\nonumber\\
&\quad
+\mathcal K_3r^3
+\mathcal K_2r^2
+\mathcal K_1r
+\mathcal K_0
+\frac{\mathcal K_{-1}}{r}
+\frac{\mathcal K_{-2}}{r^2} + \dots
\biggr).
\end{align}

\begin{align}
\mathcal K_{10}
={}&
252\left(\beta_0\tilde{\mathcal F}_2\omega^{\mathrm K}_{n2}
+10\mathcal E_3\right)\Sigma_7,
\\
\mathcal K_9
={}&
\frac{56}{E}
\biggl[
\beta_0\omega^{\mathrm K}_{n2}
\biggl(
5E\tilde{\mathcal F}_1\Sigma_7
+\tilde{\mathcal F}_2
\left(5E\Sigma_6+\frac{9}{2}\alpha\Sigma_7\right)
\biggr)
\nonumber\\
&\quad
+45E
\left(\mathcal E_3\Sigma_6+\mathcal E_2\Sigma_7\right)
\biggr],
\\
\mathcal K_8
={}&
\frac{7}{E}
\biggl[
360E
\left(\mathcal E_3\Sigma_5+\mathcal E_2\Sigma_6+\mathcal E_1\Sigma_7\right)
\nonumber\\
&\quad
+\beta_0
\biggl(
5\omega^{\mathrm K}_{n2}
\bigl[
\tilde{\mathcal F}_2(9E\Sigma_5+8\alpha\Sigma_6)
+\tilde{\mathcal F}_1(9E\Sigma_6+8\alpha\Sigma_7)
+9E\tilde{\mathcal F}_0\Sigma_7
\bigr]
\nonumber\\
&\quad
-72 L_{\mathrm K}\tilde{\mathcal F}_2\Sigma_7
\biggr)
\biggr],
\\
\mathcal K_7
={}&
\frac{5}{E}
\biggl[
504E
\left(\mathcal E_3\Sigma_4+\mathcal E_2\Sigma_5+\mathcal E_1\Sigma_6
+\mathcal E_0\Sigma_7\right)
\nonumber\\
&\quad
+\beta_0
\biggl(
9\omega^{\mathrm K}_{n2}
\bigl[
\tilde{\mathcal F}_2(8E\Sigma_4+7\alpha\Sigma_5)
+\tilde{\mathcal F}_1(8E\Sigma_5+7\alpha\Sigma_6)
\nonumber\\
&\quad
+\tilde{\mathcal F}_0(8E\Sigma_6+7\alpha\Sigma_7)
+8E\tilde{\mathcal F}_{-1}\Sigma_7
\bigr]
-112 L_{\mathrm K}
(\tilde{\mathcal F}_2\Sigma_6+\tilde{\mathcal F}_1\Sigma_7)
\biggr)
\biggr],\\
\mathcal K_6
={}&
\frac{15}{E}
\biggl[
168E
\left(
\mathcal E_3\Sigma_3+\mathcal E_2\Sigma_4+\mathcal E_1\Sigma_5
+\mathcal E_0\Sigma_6+\mathcal E_{-1}\Sigma_7
\right)
\nonumber\\
&\quad
+\beta_0
\biggl(
4\omega^{\mathrm K}_{n2}
\bigl[
\tilde{\mathcal F}_2(7E\Sigma_3+6\alpha\Sigma_4)
+\tilde{\mathcal F}_1(7E\Sigma_4+6\alpha\Sigma_5)
\nonumber\\
&\quad
+\tilde{\mathcal F}_0(7E\Sigma_5+6\alpha\Sigma_6)
+\tilde{\mathcal F}_{-1}(7E\Sigma_6+6\alpha\Sigma_7)
\bigr]
\nonumber\\
&\quad
-42 L_{\mathrm K}
(\tilde{\mathcal F}_2\Sigma_5+\tilde{\mathcal F}_1\Sigma_6+\tilde{\mathcal F}_0\Sigma_7)
\biggr)
\biggr],
\\
\mathcal K_5
={}&
\frac{12}{E}
\biggl[
210E
\left(
\mathcal E_3\Sigma_2+\mathcal E_2\Sigma_3+\mathcal E_1\Sigma_4
+\mathcal E_0\Sigma_5+\mathcal E_{-1}\Sigma_6+\mathcal E_{-2}\Sigma_7
\right)
\nonumber\\
&\quad
+\beta_0
\biggl(
7\omega^{\mathrm K}_{n2}
\bigl[
\tilde{\mathcal F}_2(6E\Sigma_2+5\alpha\Sigma_3)
+\tilde{\mathcal F}_1(6E\Sigma_3+5\alpha\Sigma_4)
\nonumber\\
&\quad
+\tilde{\mathcal F}_0(6E\Sigma_4+5\alpha\Sigma_5)
+\tilde{\mathcal F}_{-1}(6E\Sigma_5+5\alpha\Sigma_6)
\bigr]
\nonumber\\
&\quad
-60 L_{\mathrm K}
(\tilde{\mathcal F}_2\Sigma_4+\tilde{\mathcal F}_1\Sigma_5
+\tilde{\mathcal F}_0\Sigma_6+\tilde{\mathcal F}_{-1}\Sigma_7)
\biggr)
\biggr],\\
\mathcal K_4
={}&
\frac{42}{E}
\biggl[
60E
\left(
\mathcal E_3\Sigma_1+\mathcal E_2\Sigma_2+\mathcal E_1\Sigma_3
+\mathcal E_0\Sigma_4+\mathcal E_{-1}\Sigma_5+\mathcal E_{-2}\Sigma_6
\right)
\nonumber\\
&\quad
+\beta_0
\biggl(
3\omega^{\mathrm K}_{n2}
\bigl[
\tilde{\mathcal F}_2(5E\Sigma_1+4\alpha\Sigma_2)
+\tilde{\mathcal F}_1(5E\Sigma_2+4\alpha\Sigma_3)
\nonumber\\
&\quad
+\tilde{\mathcal F}_0(5E\Sigma_3+4\alpha\Sigma_4)
+\tilde{\mathcal F}_{-1}(5E\Sigma_4+4\alpha\Sigma_5)
\bigr]
\nonumber\\
&\quad
-20 L_{\mathrm K}
(\tilde{\mathcal F}_2\Sigma_3+\tilde{\mathcal F}_1\Sigma_4
+\tilde{\mathcal F}_0\Sigma_5+\tilde{\mathcal F}_{-1}\Sigma_6)
\biggr)
\biggr],
\\
\mathcal K_3
={}&
\frac{42}{E}
\biggl[
60E
\left(
\mathcal E_3\Sigma_0+\mathcal E_2\Sigma_1+\mathcal E_1\Sigma_2
+\mathcal E_0\Sigma_3+\mathcal E_{-1}\Sigma_4+\mathcal E_{-2}\Sigma_5
\right)
\nonumber\\
&\quad
+\beta_0
\biggl(
5\omega^{\mathrm K}_{n2}
\bigl[
\tilde{\mathcal F}_2(4E\Sigma_0+3\alpha\Sigma_1)
+\tilde{\mathcal F}_1(4E\Sigma_1+3\alpha\Sigma_2)
\nonumber\\
&\quad
+\tilde{\mathcal F}_0(4E\Sigma_2+3\alpha\Sigma_3)
+\tilde{\mathcal F}_{-1}(4E\Sigma_3+3\alpha\Sigma_4)
\bigr]
\nonumber\\
&\quad
-24 L_{\mathrm K}
(\tilde{\mathcal F}_2\Sigma_2+\tilde{\mathcal F}_1\Sigma_3
+\tilde{\mathcal F}_0\Sigma_4+\tilde{\mathcal F}_{-1}\Sigma_5)
\biggr)
\biggr],\\
\mathcal K_2
={}&
\frac{210}{E}
\biggl[
12E
\left(
\mathcal E_3\Sigma_{-1}+\mathcal E_2\Sigma_0+\mathcal E_1\Sigma_1
+\mathcal E_0\Sigma_2+\mathcal E_{-1}\Sigma_3+\mathcal E_{-2}\Sigma_4
\right)
\nonumber\\
&\quad
+\beta_0
\biggl(
\omega^{\mathrm K}_{n2}
\bigl[
\tilde{\mathcal F}_2(6E\Sigma_{-1}+4\alpha\Sigma_0)
+\tilde{\mathcal F}_1(6E\Sigma_0+4\alpha\Sigma_1)
\nonumber\\
&\quad
+\tilde{\mathcal F}_0(6E\Sigma_1+4\alpha\Sigma_2)
+\tilde{\mathcal F}_{-1}(6E\Sigma_2+4\alpha\Sigma_3)
\bigr]
\nonumber\\
&\quad
-6 L_{\mathrm K}
(\tilde{\mathcal F}_2\Sigma_1+\tilde{\mathcal F}_1\Sigma_2
+\tilde{\mathcal F}_0\Sigma_3+\tilde{\mathcal F}_{-1}\Sigma_4)
\biggr)
\biggr],
\\
\mathcal K_1
={}&
420
\biggl[
6\left(
\mathcal E_3\Sigma_{-2}+\mathcal E_2\Sigma_{-1}+\mathcal E_1\Sigma_0
+\mathcal E_0\Sigma_1+\mathcal E_{-1}\Sigma_2+\mathcal E_{-2}\Sigma_3
\right)
\nonumber\\
&\quad
+\frac{\beta_0}{E}
\biggl(
3\omega^{\mathrm K}_{n2}
\bigl[
\tilde{\mathcal F}_2(2E\Sigma_{-2}+\alpha\Sigma_{-1})
+\tilde{\mathcal F}_1(2E\Sigma_{-1}+\alpha\Sigma_0)
\nonumber\\
&\quad
+\tilde{\mathcal F}_0(2E\Sigma_0+\alpha\Sigma_1)
+\tilde{\mathcal F}_{-1}(2E\Sigma_1+\alpha\Sigma_2)
\bigr]
\nonumber\\
&\quad
-4 L_{\mathrm K}
(\tilde{\mathcal F}_2\Sigma_0+\tilde{\mathcal F}_1\Sigma_1
+\tilde{\mathcal F}_0\Sigma_2+\tilde{\mathcal F}_{-1}\Sigma_3)
\biggr)
\nonumber\\
&\quad
+\frac{12\beta_0q_e^2\omega^{\mathrm K}_{n2}}
{p(q_e+1)}
(\tilde{\mathcal F}_2\Sigma_{-3}+\tilde{\mathcal F}_1\Sigma_{-2}
+\tilde{\mathcal F}_0\Sigma_{-1}+\tilde{\mathcal F}_{-1}\Sigma_0)
\biggr],\\
\mathcal K_0
={}&
1260
\biggl[
2\left(
\mathcal E_3\Sigma_{-3}+\mathcal E_2\Sigma_{-2}+\mathcal E_1\Sigma_{-1}
+\mathcal E_0\Sigma_0+\mathcal E_{-1}\Sigma_1+\mathcal E_{-2}\Sigma_2
\right)
\nonumber\\
&\quad
+\frac{\beta_0}{Ep(q_e+1)}
\biggl(
2\omega^{\mathrm K}_{n2}
\biggl[
\tilde{\mathcal F}_2
\bigl(
[2\alpha q_e^2+Ep((q_e+1)\ell_e-2)]\Sigma_{-3}
+p\alpha(q_e+1)\Sigma_{-2}
\bigr)
\nonumber\\
&\quad
+\tilde{\mathcal F}_1
\bigl(
[2\alpha q_e^2+Ep((q_e+1)\ell_e-2)]\Sigma_{-2}
+p\alpha(q_e+1)\Sigma_{-1}
\bigr)
\nonumber\\
&\quad
+\tilde{\mathcal F}_0
\bigl(
[2\alpha q_e^2+Ep((q_e+1)\ell_e-2)]\Sigma_{-1}
+p\alpha(q_e+1)\Sigma_0
\bigr)
\nonumber\\
&\quad
+\tilde{\mathcal F}_{-1}
\bigl(
[2\alpha q_e^2+Ep((q_e+1)\ell_e-2)]\Sigma_0
+p\alpha(q_e+1)\Sigma_1
\bigr)
\biggr]
\nonumber\\
&\quad
-2 p L_{\mathrm K}(q_e+1)
(\tilde{\mathcal F}_2\Sigma_{-1}+\tilde{\mathcal F}_1\Sigma_0
+\tilde{\mathcal F}_0\Sigma_1+\tilde{\mathcal F}_{-1}\Sigma_2)
\biggr)
\biggr],\\
\mathcal K_{-1}
={}&
2520
\biggl[
\mathcal E_3\Sigma_{-4}+\mathcal E_2\Sigma_{-3}+\mathcal E_1\Sigma_{-2}
+\mathcal E_0\Sigma_{-1}+\mathcal E_{-1}\Sigma_0+\mathcal E_{-2}\Sigma_1
\nonumber\\
&\quad
-\frac{\beta_0}{Ep(q_e+1)}
\biggl(
p\omega^{\mathrm K}_{n2}
\biggl[
\tilde{\mathcal F}_2
\bigl(E(q_e+1)\Sigma_{-4}
-\alpha((q_e+1)\ell_e-2)\Sigma_{-3}\bigr)
\nonumber\\
&\quad
+\tilde{\mathcal F}_1
\bigl(E(q_e+1)\Sigma_{-3}
-\alpha((q_e+1)\ell_e-2)\Sigma_{-2}\bigr)
\nonumber\\
&\quad
+\tilde{\mathcal F}_0
\bigl(E(q_e+1)\Sigma_{-2}
-\alpha((q_e+1)\ell_e-2)\Sigma_{-1}\bigr)
\nonumber\\
&\quad
+\tilde{\mathcal F}_{-1}
\bigl(E(q_e+1)\Sigma_{-1}
-\alpha((q_e+1)\ell_e-2)\Sigma_0\bigr)
\biggr]
\nonumber\\
&\quad
+ 2 L_{\mathrm K}
\biggl[
2q_e^2
(\tilde{\mathcal F}_2\Sigma_{-3}+\tilde{\mathcal F}_1\Sigma_{-2}
+\tilde{\mathcal F}_0\Sigma_{-1}+\tilde{\mathcal F}_{-1}\Sigma_0)
\nonumber\\
&\quad
+p(q_e+1)
(\tilde{\mathcal F}_2\Sigma_{-2}+\tilde{\mathcal F}_1\Sigma_{-1}
+\tilde{\mathcal F}_0\Sigma_0+\tilde{\mathcal F}_{-1}\Sigma_1)
\biggr]
\biggr)
\biggr],
\\
\mathcal K_{-2}
={}&
2520
\left(
\mathcal E_3\Sigma_{-5}+\mathcal E_2\Sigma_{-4}
+\mathcal E_1\Sigma_{-3}+\mathcal E_0\Sigma_{-2}
+\mathcal E_{-1}\Sigma_{-1}+\mathcal E_{-2}\Sigma_0
\right)
\nonumber\\
&\quad
-\frac{1260\beta_0}{E(q_e+1)}
\biggl[
(q_e+1)\omega^{\mathrm K}_{n2}
\bigl[
\tilde{\mathcal F}_2(E\Sigma_{-5}+2\alpha\Sigma_{-4})
+\tilde{\mathcal F}_1(E\Sigma_{-4}+2\alpha\Sigma_{-3})
\nonumber\\
&\quad
+\tilde{\mathcal F}_0(E\Sigma_{-3}+2\alpha\Sigma_{-2})
+\tilde{\mathcal F}_{-1}(E\Sigma_{-2}+2\alpha\Sigma_{-1})
\bigr]
\nonumber\\
&\quad
+4 L_{\mathrm K}((q_e+1)\ell_e-2)
(\tilde{\mathcal F}_2\Sigma_{-3}+\tilde{\mathcal F}_1\Sigma_{-2}
+\tilde{\mathcal F}_0\Sigma_{-1}+\tilde{\mathcal F}_{-1}\Sigma_0)
\biggr].
\end{align}
\end{subequations}
With this, $Z^+_{n,\ell=2,m=2}$ is obtained by
\begin{equation}
    Z_{n,\ell=2,m=2}^+ = \frac{C_{\rm {trans}}\sqrt{\pi/5}\,\omega^4}
{7620480\,W_r\,E} \sum_{j=-2}^{10}\mathcal{K}_j \times \left\{\frac{r^j}{1+\frac{\alpha}{E r}}\right\}^{n2}_{\mathrm K}
\end{equation}
where $\{\}_{\mathrm K}$ is given in~\eqref{KepElem} of the main text.